\documentclass[10pt,aps,prd,twocolumn,nofootinbib,amssymb,preprintnumbers,superscriptaddress]{revtex4-2}
\usepackage{graphicx}
\usepackage[justification=justified,singlelinecheck=false]{caption}
\usepackage{caption}
\usepackage{subcaption}
\usepackage{dcolumn}
\usepackage{bm}
\usepackage{multirow}
\usepackage[dvipsnames]{xcolor}
\usepackage{color}
\usepackage{ragged2e}
\usepackage{pifont}
\usepackage{bbding}
\usepackage{comment}
\usepackage{dirtytalk}
\usepackage{comment}
\usepackage{verbatim}
\usepackage{amsmath}
\usepackage{amssymb}
\usepackage{amsfonts}
\usepackage{cases}
\usepackage{tensor}
\usepackage{appendix}
\usepackage{tabularx}

\newcommand{\til}{~}
\usepackage[T1]{fontenc} 
\usepackage{tikz, pgfplots}
\pgfplotsset{compat=1.18}
\usetikzlibrary{positioning}
\usetikzlibrary{fadings}

\tikzfading 
[
  name=fade out,
  inner color=transparent!0,
  outer color=transparent!100
]

\definecolor{grey}{rgb}{0.4,0.4,0.4}
\definecolor{dullmagenta}{rgb}{0.4,0,0.4}
\definecolor{darkblue}{rgb}{0,0,0.4}
\definecolor{midblue}{rgb}{0,0,0.5}
\definecolor{midred}{rgb}{0.5,0,0}
\definecolor{orange}{rgb}{1,0.5,0}
\definecolor{lightbrown}{rgb}{0.75,0.5,0.25}
\definecolor{tan}{cmyk}{0.14,0.42,0.56,0}
\definecolor{djunglegreen}{cmyk}{0.99,0,0.52,0}
\definecolor{lightgreen}{rgb}{0,1,0}
\definecolor{olivegreen}{cmyk}{0.64,0,0.95,0.40}
\definecolor{midgreen}{rgb}{0.0,0.675,0.0}
\definecolor{darkgreen}{rgb}{0,0.5,0}
\definecolor{ceruleanblue}{rgb}{0.0, 0.45, 1} 
\definecolor{cyan}{HTML}{0ABAB5}
\definecolor{burgundy}{rgb}{0.5, 0.0, 0.13}
\definecolor{hvred}{RGB}{186,12,47}
\definecolor{ste}{rgb}{0.01, 0.28, 1.0}
\usepackage[colorlinks=true, citecolor=ceruleanblue, filecolor=ceruleanblue, linkcolor=ceruleanblue, urlcolor=ceruleanblue]{hyperref}

\newcolumntype{Y}{>{\centering\arraybackslash}X}

\begin{document}
\title{Gravitational-wave dispersion over inhomogeneous space-times:\\ General relativity, screened theories of gravity and non-minimal dark energy}

\author{Nicola Menadeo}
\email{nicola.menadeo@aei.mpg.de}
\affiliation{Scuola Superiore Meridionale, Largo San Marcellino 10, I-80138, Napoli, Italy}
\affiliation{Max Planck Institute for Gravitational Physics (Albert Einstein Institute), \\
Am Mühlenberg 1, D-14476 Potsdam-Golm, Germany}

\author{Serena Giardino}
\email{s.giardino@imperial.ac.uk}
\affiliation{Max Planck Institute for Gravitational Physics (Albert Einstein Institute), \\
Am Mühlenberg 1, D-14476 Potsdam-Golm, Germany}
\affiliation{Abdus Salam Centre for Theoretical Physics, Imperial College, London, SW7 2AZ, UK}

\author{Miguel Zumalac\'arregui}
\email{miguel.zumalacarregui@aei.mpg.de}
\affiliation{Max Planck Institute for Gravitational Physics (Albert Einstein Institute), \\
Am Mühlenberg 1, D-14476 Potsdam-Golm, Germany}

\begin{abstract}
Gravitational waves (GWs) are direct probes of cosmological gravity, sensitive to space-time inhomogeneities along their propagation. The presence of massive objects breaks homogeneity and isotropy, allowing for new interactions between different GW polarizations, and opening up the intriguing opportunity to test modified gravity theories. This setup generalizes the notion of gravitational deflection and lensing, revealing novel phenomena in modified theories.  Any non-minimal theory introduces effective mass terms for GWs, causing \textit{lens-induced dispersion} (LID), a frequency-dependent phase shift on the waveform. 
We compute GW dispersion in Einstein's general relativity (GR) for a spherical matter distribution, finding a small but non-zero phasing that is potentially accessible to next-generation detectors. 
We then extend our analysis to scalar-tensor theories, focusing on symmetron gravity as an example of screened theory, combining cosmological deviations and consistency with local gravity tests. We find enhanced GW dispersion in a large region of the symmetron parameter space, compared to both GR and Brans-Dicke theory.
We argue that dispersion, associated to an effective mass for the metric fluctuations, can in some cases prevent the propagation of GWs through some astrophysical bodies, turning them into reflectors. 
Our analysis shows that the Earth becomes an efficient GW shield for a hitherto unconstrained region of the symmetron parameter space, leading to a $\sim 50\%$ fraction of events becoming unobservable or at least displaying a dramatic modification of the detector antenna response.
The richness and universality of dispersive phenomena in non-minimal theories open a new avenue to test theories of dynamical dark-energy, relevant in light of recent observational results challenging the $\Lambda$CDM paradigm.

\end{abstract}

\maketitle
{\hypersetup{hidelinks}
  \tableofcontents}

\section{Introduction}
The discovery of GWs by the LIGO–Virgo–KAGRA (LVK) collaboration has transformed our view of the universe, granting new tools to test gravity in highly dynamical and strong-field settings. Notable events like GW150914 \cite{LIGOScientific:2016aoc} (the first detection of a black hole merger) and GW170817 \cite{Abbott_2017_GW170817} (the first multi-messenger event, a neutron star binary merger with an electromagnetic counterpart) have pushed tests of General Relativity (GR) into regimes previously unattainable \cite{LIGOScientific:2019fpa,LIGOScientific:2020tif,LIGOScientific:2021sio}.

Yet, relying on GR to describe the evolution of the universe within $\Lambda$CDM cosmology still requires the existence of mysterious dark matter \cite{Cirelli:2024ssz,Taoso:2007qk,Bertone:2004pz} and dark energy \cite{SupernovaSearchTeam:1998fmf,Li:2011sd,Frieman:2008sn} to explain observations. The fundamental nature of these components remains elusive and constitutes one of the most fundamental puzzles in physics \cite{Huterer:2017buf,Cortes:2024lgw,Buckley:2017ijx,Hogan:2000bv,Spergel:1999mh,Goodman:1984dc}. A compelling alternative to the presence of dark energy is the hypothesis that cosmic acceleration arises from a modification of gravity on large scales (see e.g. \cite{AmendolaTsujikawabook, Oks:2021hef, Clifton:2011jh} for reviews). This idea is powerful because the presence of dark energy (and dark matter) is only indirectly inferred through gravitational effects, which are inevitably affected by our ignorance of gravity's behavior on extragalactic scales (where GR has not been as strongly constrained as within our Solar System) \cite{Planck:2015bue,Hu:2024wub,Thomas:2025qiz,Ishak:2018his}. This has led to the development of many alternative gravity theories \cite{Clifton:2011jh,Capozziello:2006ra,Shankaranarayanan:2022wbx,Joyce:2014kja}, especially scalar‑tensor models (with a large class of theories unified under Horndeski's Lagrangian \cite{Horndeski:2024sjk, Kobayashi:2019hrl,Horndeski:2024sjk} and its extensions \cite{Zumalacarregui:2013pma,Gleyzes:2014dya,BenAchour:2016fzp}), which introduce additional light scalar fields. These new fields can mediate a fifth force that alters cosmic dynamics, offering potential explanations for phenomena typically attributed to the dark sector. Importantly, many of these alternative models can now be strongly constrained by GW observations and possibly ruled out. The most impactful of these observations so far has been the coincident detection of GW170817 and its electromagnetic counterpart GRB 170817A \cite{LIGOScientific:2017zic}, which placed an extremely tight bound on the GW propagation speed \cite{LIGOScientific:2017zic}
\begin{equation}
-3\cdot10^{-15}\leq\dfrac{c_{\rm GW}-c}{c}\leq +7\cdot10^{-16},
\end{equation}
ruling out many dark energy scenarios \cite{Ezquiaga:2017ekz, Bettoni:2016mij, Baker:2017hug} (although see \cite{deRham:2018red} for an important caveat).

In addition to emission tests, GW propagation provides a complementary and independent probe of gravity. Standard analyses in propagation typically assume a homogeneous and isotropic cosmological background, described by the Friedmann–Lema\^{i}tre–Robertson–Walker (FLRW) metric \cite{Ezquiaga:2018btd, Bian:2025ifp,Ezquiaga:2021ler}. In this idealized setting, scalar, vector, and tensor perturbations decouple at linear order, severely limiting the range of observable phenomena \cite{Malik:2008im}. However, the universe is of course not perfectly homogeneous: structures such as galaxies, galaxy clusters, and even small-scale astrophysical objects such as stars and planets introduce inhomogeneities that act as gravitational lenses \cite{Perlick:2004tq,1967ApJ...150..737G,Seitz:1994xf}. These inhomogeneities break the high degree of symmetry of the FLRW metric, allowing for mixing and non-trivial interactions between the standard tensor polarizations ($+$, $\times$) and any additional gravitational degrees of freedom \cite{Ezquiaga:2020dao,Cayuso:2024ppe,Streibert:2024cuf,Takeda:2024ghe,Diedrichs:2023foj}.

Unlike light, which is frequently scattered, absorbed, or delayed by intervening matter, GWs propagate essentially undisturbed across cosmological distances. This remarkable transparency of the universe to GWs makes them pristine messengers of the gravitational interaction, opening a new window on the universe at novel frequencies that can provide precious information about regions opaque to light. In general, gravitational lensing alters GW signals through both deflection and magnification.
In the case of strong lensing, the signal can split into multiple images, appearing as distinct echoes in the observed waveform \cite{Xu:2021bfn,Xin:2021zir,2017CQGra..34v5005P,Janquart:2023mvf,LIGOScientific:2023bwz}.
The primary observable signature of GW lensing is typically the time delay between these images, or between different polarizations (in modified gravity).
While current GW detectors have limited source localization, they possess great time resolution: With the, and even more so with next-generation observatories like LISA, lensed GW events promise to provide powerful tests of gravity. Specifically, in the  upcoming LVK O5 run we expect $1-10$/yr gravitationally magnified stellar mass black holes \cite{Smith:2022vbp} and $\sim 1$/yr split into multiple observable images \cite{Wierda:2021upe}, with order-of-magnitude increases with the Einstein Telescope and Cosmic Explorer \cite{Li:2018prc,Jana:2024uta}. The low-frequency space-borne detector LISA will be very sensitive to multiply-imaged massive black hole mergers \cite{Sereno:2010dr,Gutierrez:2025ymd} and signal distortions by cosmic structures \cite{Caliskan:2023zqm,Brando:2024inp,Singh:2025uvp}.
GW lensing allows for several applications: e.g., lensing magnification can help discover otherwise inaccessible distant astrophysical sources \cite{Lo:2024wqm}.
Additionally, GW lensing is sensitive to dark matter (sub-)structure in the lens, which is often a galaxy, providing important information on dark matter properties \cite{Tambalo:2022wlm, Singh:2025uvp, Cheung:2024ugg,Zumalacarregui:2024ocb}.

In this work, we are interested in the potential of GW lensing to test modified gravity and provide constraints on such alternative theories.
This is possible because distinct classes of modified gravity theories imprint characteristically different signatures on the waveform. As summarized in Figure 1 of \cite{Menadeo:2024uoq}, theories involving two derivatives of the tensor and scalar fields exhibit birefringence, which provides strong constraints, comparable in some limits to those given by the multi-messenger event \cite{Goyal:2023uvm} (see also  \cite{Serra:2022pzl}). Theories involving one derivative show frequency oscillations \cite{Garoffolo:2019mna}. In this work we focus on dispersion, appearing as frequency-dependent modifications of the phase, which is universal in all modified theories that include even a simple mass term (see \cite{deRham:2016nuf} for the case of massive gravity).
In general, GW lensing provides a plethora of observational signatures (such as waveform distortions, magnifications, time delays, or frequency-dependent effects) that do not require electromagnetic counterparts and can be observed across a broad range of astrophysical events. Dispersive effects are generally weaker than those of birefringence, for instance, but they occur for all modified theories instead of for a restricted class containing two derivative of the fields. Therefore, GW lensing has the potential to test both the propagation properties of GWs and the underlying theory of gravity itself \cite{Oancea:2020khc,Narola:2023viz}.

In the classical high-frequency regime, where the GW wavelength $\lambda$ is much smaller than the curvature radius of the lens (e.g.\ its Schwarzschild radius $R_s$), the geometric optics (GO) approximation holds, and GWs propagate along null geodesics with parallel transported polarization \cite{Fleury:2015hgz,PhysRev.166.1263,Cabral:2016klm,Dolan:2017zgu}. Historically, most analyses of GW lensing (in GR or beyond) adopt this limit \cite{Andersson:2020gsj,Kubota:2022lbn,Hou:2019wdg,Koksbang:2021alx,Garoffolo:2019mna}. However, for low-frequency GWs (such as those relevant for Pulsar Timing Arrays (PTAs) or the space-based detector LISA) the condition $\lambda \ll R_s$ can be easily violated. In these regimes, the GW wavelength may become comparable to the characteristic size of astrophysical structures, and the presence of wave optics features becomes unavoidable. Past detections involved high-frequency waves for which GO remained firmly valid, but with the advent of new observational windows, we need to take into account corrections to this approximation, with the long-term goal of developing a full wave-optics treatment of GW lensing.

The resulting corrections to GO, collectively referred to as \emph{beyond geometric optics} (bGO) \cite{Anile:1976gq,Harte:2018wni,Dolan:2017zgu,Harte:2019tid}, introduce frequency-dependent modifications to the GW amplitude and phase, such as in \emph{lens-induced dispersion} (LID) \cite{Oancea:2020khc,Oancea:2022szu,Menadeo:2024uoq}. 

The authors of \cite{Cusin:2017fwz} carried out the first investigations into bGO corrections to GW propagation within GR, introducing the necessary theoretical framework and considering a simple point mass lens \cite{Cusin:2019rmt}. It was shown that, since in the bGO regime the polarization tensor is no longer constrained to be transverse to the propagation direction, an apparent longitudinal component emerges. This effect is of geometric origin and may be interpreted as a mild rotation or distortion of the polarization plane.
This framework was later extended to the simplest scalar-tensor theory, Brans-Dicke, in \cite{Menadeo:2024uoq}, which analyzed the interplay between scalar and tensor radiation (at the first order bGO) and introduced the notion of LID. In this setting, it was found that, beyond the emergence of apparent polarizations, bGO effects also induce genuine modifications of the physical $+$ and $\times$ polarizations, opening a novel avenue for testing gravity beyond GR.

To satisfy the stringent observational constraints in the Solar System, all viable modified gravity theories must invoke screening mechanisms, which suppress deviations from GR in high-density regions (such as the Solar System itself) while allowing them in lower-density environments \cite{terHaar:2020xxb,Brax:2013ida,Olive:2007aj}. In the latter, at large cosmological scales, modified gravity effects might explain the accelerated expansion of the universe without dark energy. There are several types of screening mechanisms, including chameleon \cite{Khoury:2003rn,Burrage:2017qrf,Khoury:2003aq}, Vainshtein \cite{Dar:2018dra,Babichev:2013usa,Crisostomi:2017lbg}, and symmetron \cite{Hinterbichler:2010es,Hinterbichler:2011ca,Burrage:2018zuj,Winther:2011qb,Pietroni:2005pv}. The symmetron mechanism, the focus of this work, operates via a $\mathbb{Z}_2$ symmetry: in high-density regions the symmetry is restored and the scalar field is suppressed, while in low-density environments it is broken and the field acquires a vacuum expectation value (VEV) that mediates a scalar force.

Symmetron gravity is motivated both from the theoretical and the observational point of view: through the lens of effective field theory, the symmetry it is endowed with can stabilize the scalar potential against quantum corrections \cite{Hinterbichler:2010es}, and similar non-minimally coupled scalar-tensor theories seem to be preferred in light of recent results hinting at dynamical dark energy \cite{DESI:2024mwx, Ye:2024ywg, Wolf:2025jed}.
Screened modified gravity theories are notoriously hard to constrain: typically, some region of the parameter space can be excluded by laboratory experiments and astrophysical and cosmological observations, but, especially in symmetron gravity, there are large swaths of parameter space which remain unconstrained \cite{Burrage:2017qrf, Brax:2021wcv, Sabulsky:2018jma, Elder:2016yxm, Elder:2019yyp}.
Screened theories have not been properly studied in the context of GW propagation, and especially beyond the geometric optics approximation: \cite{Ezquiaga:2020dao} studied the effect of Vainshtein screening in quartic Galileon theories, but only within the GO regime. Moreover, symmetron gravity and similar non-minimally coupled models (exhibiting $G_{2,XX}=0$ in Horndeski language) predict luminal scalar waves (propagating at $c$), which makes their treatment easier, since both tensor and scalar waves share the same lightcone structure.

In this work, we go beyond these limitations, investigating GW propagation beyond geometric optics through a spherically symmetric, extended lens, in both GR and symmetron gravity. Employing the formalism from \cite{Menadeo:2024uoq}, we compute the dispersive corrections arising from a constant density matter profile. We show that even in GR, extended lenses introduce nontrivial dispersive effects on the standard GW polarizations, in contrast to the point-mass case. Matter fields produce small GW dispersion even in Einstein's theory. 
Moreover, in the symmetron scenario, the interaction between the GW and the scalar field (whose spatial profile depends on the lens’s density) significantly amplifies frequency-dependent corrections. 
LID can be understood as $h_+,h_\times$ acquiring an effective mass, whose value depends on the background scalar field. We also show how this mass can become large enough to prevent the propagation of GWs through matter: GWs are blocked by Earth in some regions of the symmetron parameter space, radically changing the way these signals are observed in these theories.

The work is organized as follows. In Sec.~\ref{General formalism for GW propagation}, we present the general formalism for GW propagation in scalar–tensor theories, highlighting how bGO corrections affect the polarization structure. In Sec.~\ref{symmetron}, we review symmetron gravity and its screening mechanism, presenting the scalar field solutions for a static, spherically symmetric body with constant density in the Jordan frame. In Sec.~\ref{homogeneous spherical lens}, we compute the dispersive corrections induced by a homogeneous spherical lens for GR and symmetron gravity. In Sec.~\ref{sec:critical_freq_constraints}, we introduce the concept of a critical GW frequency, at which the propagation of GWs is effectively blocked by extreme dispersion analogous to a plasma frequency for electrons. We also provide the constraints on the symmetron parameters that arise from GW dispersion and ideally complement those from other experiments. We conclude in Sec.~\ref{Conclusion} and outline future developments. In the following, we use $\hbar=c=1$.

\section{General formalism for GW propagation}\label{General formalism for GW propagation}

Scalar-tensor theories of gravity feature both metric and scalar degrees of freedom. In addition to the usual spin-2 field $g_{\mu\nu}$, a dynamical scalar field $\phi$ is present, introducing an extra mode absent in GR. Perturbing the field equations around a background solution (labeled with a bar on the dynamic variables) $\{g_{\mu\nu}, \phi\}$ generally gives rise to two types of wave-like excitations: tensor waves, associated with perturbations of the metric, and scalar waves, arising from fluctuations of the scalar field. Throughout this work, only tensor perturbations will be considered; scalar waves are neglected, and the scalar field enters in the dynamics through its background configuration $\phi$ (see Sec.~\ref{sec:Scalar field solutions}). This assumption is motivated by the fact that this work focuses on scalar-tensor theories featuring screening mechanisms, where the emission of scalar radiation is often suppressed under realistic astrophysical conditions (see, e.g.,  \cite{Garoffolo:2019mna} for chameleon and symmetron models, and  \cite{Dar:2018dra} for the Vainshtein mechanism). Such suppression is highly sensitive to the specific physical configuration of the gravitational system (such as coupling strengths in the scalar-tensor Lagrangian and environmental conditions) and cannot be assumed \textit{a priori}. Nevertheless, neglecting scalar waves remains a justified and convenient approximation for the purposes of the present analysis. For a treatment that includes scalar waves in a similar context, see  \cite{Menadeo:2024uoq,Dalang:2020eaj}.

We start by introducing the metric perturbation
\begin{equation}\label{metric perturbation}
    g_{\mu\nu}^{\text{tot}}\equiv g_{\mu\nu}+h_{\mu\nu},
\end{equation}
where $g_{\mu\nu}$ is the background metric and $h_{\mu\nu}$ the metric perturbation. We also introduce
\begin{equation}
    \phi^{\rm tot}\equiv \phi+\delta\phi,
\end{equation}
where $\phi$ is the background scalar field and $\delta\phi$ its perturbation, which we will neglect in the following.
The GW propagation equation can be derived in terms of the well-known trace-reversed metric perturbation $\Tilde{h}_{\mu\nu}$, by employing (when feasible) the transverse and traceless conditions (or equivalently TT gauge)
\begin{align}\label{eq: trace-reversed metric}
    \tilde{h}_{\mu\nu}&\equiv h_{\mu\nu}-\frac{1}{2}g_{\mu\nu}h,\\ \label{transverse condition}
\nabla^{\mu}\tilde{h}_{\mu\nu}&=0,\\
\label{traceless condition}
\hspace{0.3cm} \tilde{h}&=0.
\end{align}

It is important to point out that the TT gauge conditions cannot always be imposed in a fully consistent way: the two conditions above generally do not remain compatible throughout the GW evolution. Broadly speaking, the transverse condition is fully general and can typically be enforced by fixing the gauge freedom associated with infinitesimal diffeomorphisms. The traceless condition, however, is more restrictive. This is not guaranteed to hold in general, particularly in curved backgrounds or in theories propagating additional degrees of freedom.

In GR, the TT gauge can be consistently imposed in vacuum and remains valid far from matter sources~\cite{Maggiore:2007ulw,Flauger:2017ged}. In the presence of matter, however, the traceless condition cannot be maintained globally (see Sec.~II of~\cite{Ezquiaga:2020dao} and~\cite{RizwanaKausar:2016zgi} for a detailed discussion).

Beyond GR, the gauge-fixing procedure becomes more subtle. The trace of the tensor perturbation typically evolves independently from the residual gauge freedom, preventing its elimination even locally. Nonetheless, in many scalar–tensor theories, TT gauge conditions can still be consistently applied to isolate the radiative degrees of freedom, provided that the energy-momentum tensor enters only at the background level and does not source the perturbation equations at the same order~\cite{Maggiore:1999wm,Lobato:2024rkb,Dalang:2020eaj,Takeda:2023wqn,Berry:2011pb}.

The evolution of the tensor degrees of freedom can be compactly written as
\begin{equation}\label{eq:propagation_equation}
    \tensor{\mathsf{P}}{_\mu_\nu^\alpha^\beta}\tilde{h}_{\alpha\beta}=0,
\end{equation}
where
\begin{equation}\label{eq:propagator_operator}
\tensor{\mathsf{P}}{_\mu_\nu^\alpha^\beta}\equiv\tensor{\mathsf{K}}{_\mu_\nu^\alpha^\beta^\gamma^\rho}\nabla_{\gamma}\nabla_{\rho}+\tensor{\mathsf{A}}{_{\mu\nu}^{\alpha\beta\gamma}}\nabla_{\gamma}+\tensor{\mathsf{M}}{_{\mu\nu}^{\alpha\beta}}
\end{equation}
is the propagation differential operator encoding derivatives at all orders. the operators $\tensor{\mathsf{K}}{_\mu_\nu^\alpha^\beta^\gamma^\rho}$, $\tensor{\mathsf{A}}{_{\mu\nu}^{\alpha\beta\gamma}}$, and $\tensor{\mathsf{M}}{_{\mu\nu}^{\alpha\beta}}$ depend on the gravity theory chosen and are functions of the background quantities $g_{\mu\nu}$ and $\phi$. We provide their explicit expression for symmetron gravity, the focus of the present work, in Appendix \ref{appendix:symmetron_functions}). (These operators act only on $\tilde{h}_{\alpha\beta}$, but we do not use a tilde on them to keep the notation simple).

\subsection{Short-wave expansion}
\label{shortwave}
In order to study the behavior of GWs in curved spacetime, it is usually convenient to make use of the short-wave (or WKB) approximation. This assumes that the GW wavelength $\lambda_{\text{GW}}$ is much smaller than the scale on which the spacetime curvature varies significantly, $L_{b}$ (See Sec.~\ref{dispersive corrections in symmetron} for how this assumption can be revisited without violating the validity of the high frequency approximation). In this framework, the perturbation field $\tilde{h}_{\mu\nu}$ is expanded as a rapidly oscillating exponential modulated by a slowly varying amplitude
\begin{equation}\label{WKB decomposition metric}
\tilde{h}_{\mu\nu} = \left( \tilde{h}^{(0)}_{\mu\nu} + \epsilon \tilde{h}^{(1)}_{\mu\nu} + \cdots \right) e^{i \theta / \epsilon},
\end{equation}
where $\epsilon\equiv 2\lambda_{\text{GW}}/L_{b}$ is a dimensionless book-keeping parameter and the quantity $\theta$ denotes the phase related to the wavevector $k_\mu \equiv \nabla_\mu \theta$, while each order $\tilde{h}^{(n)}_{\mu\nu}$ in the expansion encodes amplitude corrections at increasing orders in $\epsilon$. 
Inserting Eq.~\eqref{WKB decomposition metric} into the covariant wave equation~\eqref{eq:propagation_equation} and collecting terms at each order in $\epsilon$ leads to a hierarchy of equations governing the propagation and evolution of GW modes.

\subsubsection{Geometric optics}

The leading ($\epsilon^{-2}$) and next-to-leading ($\epsilon^{-1}$) orders in the $\tilde{h}_{\mu\nu}$ expansion are referred to the geometric optics regime: in this limit, GWs propagate along null geodesics of the background spacetime and the polarization remains transverse with respect to the propagation direction. The formalism closely mirrors that of electromagnetic waves in curved space, where the phase follows the null trajectories and the amplitude evolves according to a transport equation \cite{Schneider:1992bmb,Fleury:2015hgz,Fleury:2020cal}.
At leading order in the geometric optics limit, indeed, one obtains
\begin{equation}
k^{\alpha}k_{\alpha}=0,
\end{equation}
stating that GWs propagate along null geodesics, as expected for massless fields.

The next-to-leading order, provides the generalized transport equation for the amplitude $\tilde{h}^{(0)}_{\mu\nu}$
\begin{equation}\label{eq:eps1}
\left[2\tensor{\mathsf{{K}}}{_\mu_\nu^\alpha^\beta^\gamma^\rho}k_{\rho}\nabla_{\gamma}+\tensor{\mathsf{{K}}}{_\mu_\nu^\alpha^\beta^\gamma^\rho}\nabla_{\rho}k_{\gamma}+\tensor{\mathsf{{A}}}{_{\mu\nu}^{\alpha\beta\gamma}}k_{\gamma}\right]\tilde{h}^{(0)}_{\alpha\beta}=0.
\end{equation}
In addition, combining the TT gauge conditions, specifically Eq.~\eqref{transverse condition}, with the WKB expansion~\eqref{WKB decomposition metric}, one gets
\begin{equation}\label{geometric_optics_transversality}
    k^\mu \tilde{h}^{(0)}_{\mu\nu} = 0,
\end{equation}
thus proving that the GWs polarizations encoded in the $\tilde{h}^{(0)}_{\mu\nu}$ definition remain transverse to the direction of propagation identified by $k^{\mu}$.
\subsubsection{Beyond geometric optics}\label{sec:beyond_geometric_optics}

At $\mathcal{O}(\epsilon^0)$, deviations to geometric optics start to emerge. These subleading terms encode effects such as dispersion and polarization mixing, which are intrinsically absent in the geometric optics description (see \cite{Menadeo:2024uoq} for more details). The corresponding equation for $\tilde{h}^{(1)}_{\alpha\beta}$ reads
\begin{equation}\label{eq:eps0}
\begin{split}
    &\left[2\tensor{\mathsf{K}}{_\mu_\nu^\alpha^\beta^\gamma^\rho}k_{\rho}\nabla_{\gamma}+\tensor{\mathsf{K}}{_\mu_\nu^\alpha^\beta^\gamma^\rho}\nabla_{\rho}k_{\gamma}+\tensor{A}{_{\mu\nu}^{\alpha\beta\gamma}}k_{\gamma}\right]\tilde{h}^{(1)}_{\alpha\beta}=\\&=i \tensor{\mathsf{P}}{_{\mu\nu}^{\alpha\beta}}\tilde{h}^{(0)}_{\alpha\beta},
\end{split}
\end{equation}
where the right-hand side acts as a source term generated by the propagation differential operator~\eqref{eq:propagator_operator} acting on the GO amplitude\footnote{All higher-order corrections obey the same evolution equation, with the substitution $\tilde{h}_{\alpha\beta}^{(1)}, \tilde{h}_{\alpha\beta}^{(0)} \rightarrow \tilde{h}_{\alpha\beta}^{(n)}, \tilde{h}_{\alpha\beta}^{(n-1)}$ at each successive order $n$.}. Further, Eq.~\eqref{transverse condition} reads 
\begin{equation}\label{eq:transverse gauge bgo}
k^{\mu}\tilde{h}^{(1)}_{\mu\nu}=i\nabla^{\mu}\tilde{h}^{(0)}_{\mu\nu},
\end{equation}
thereby showing that the assumption that polarizations remain orthogonal to the wavevector (as shown in Eq.~\eqref{geometric_optics_transversality}) no longer holds once subleading corrections to geometric optics are included: higher order corrections in the short-wave expansion generically induce longitudinal components in the polarization structure, an effect that arises not only for tensorial fields like gravity, but also for vector and scalar fields subject to wave propagation in curved backgrounds, as discussed in  \cite{Harte:2018wni,Harte:2019tid}.

\subsection{Tetrad decomposition}
At each order $n$ in the short-wave expansion, the amplitude $\tilde{h}^{(n)}_{\mu\nu}$ can be decomposed by means of a null tetrad basis
\begin{equation}
e^{\mu}_A\equiv\{k^\mu, m^\mu, l^\mu, n^\mu\}.
\end{equation}
The tetrads are parallel transported along the null geodesic defined by $k^\mu$, and satisfy
\begin{equation}\label{normalization tetrads}
l^\mu = \bar{m}^\mu, \qquad g_{\mu\nu} m^\mu l^\nu = -g_{\mu\nu} k^\mu n^\nu = 1,
\end{equation}
with all other inner products vanishing. The dual basis vectors are defined as
\begin{equation}
\hat{k}^\mu = -n^\mu, \quad \hat{n}^\mu = -k^\mu, \quad \hat{m}^\mu = l^\mu, \quad \hat{l}^\mu = m^\mu,
\end{equation}
and the background metric takes the form
\begin{equation}
g_{\mu\nu} = 2m_{(\mu}l_{\nu)} - 2n_{(\mu}k_{\nu)}.
\end{equation}
It is convenient to split each tensor amplitude component as follows
\begin{equation}\label{eq:tetrad_decomposition_amplitude}
\tilde{h}^{(n)}_{\mu\nu} = \tilde{\alpha}^{(n)}_{AB} \, \Theta^{AB}_{\mu\nu}, \quad \Theta^{AB}_{\mu\nu} \equiv \tfrac{1}{2}(e^A_\mu e^B_\nu + e^A_\nu e^B_\mu),
\end{equation}
where $\tilde{\alpha}^{(n)}_{AB}$ are complex coefficients and $\Theta^{AB}_{\mu\nu}$ are the polarization tensors constructed from the basis vectors. 

\subsubsection{Geometric optics}
The transversality condition~\eqref{geometric_optics_transversality}, along with~\eqref{normalization tetrads}, removes several components
\begin{equation}
\tilde{\alpha}^{(0)}_{nk} = \tilde{\alpha}^{(0)}_{nn} = \tilde{\alpha}^{(0)}_{nm} = \tilde{\alpha}^{(0)}_{nl} = 0,
\end{equation}
while the traceless condition enforces
\begin{equation}
\tilde{\alpha}^{(0)}_{ml} = 0.
\end{equation}

Further, the residual gauge freedom allows for the transformation
\begin{equation}
\tilde{h}_{\mu\nu} \rightarrow \tilde{h}_{\mu\nu} + 2 C_{(\mu} k_{\nu)},
\end{equation}
where \(C_\mu\) is an arbitrary complex vector orthogonal to \(k^\mu\). This is satisfied by $n^\mu \tilde{h}^{(0)}_{\mu\nu} = 0$, which then sets
\begin{equation}
\tilde{\alpha}^{(0)}_{kk} = \tilde{\alpha}^{(0)}_{km} = \tilde{\alpha}^{(0)}_{kl} = 0.
\end{equation}

As a result, only the physical helicity components survive
\begin{equation}\label{helicity geometric optics}
\tilde{h}^{(0)}_{\mu\nu} = \tilde{\alpha}^{(0)}_{mm} m_\mu m_\nu + \tilde{\alpha}^{(0)}_{ll} l_\mu l_\nu
\end{equation}
and can be directly related to the usual GW polarizations $h_+$ and $h_{\times}$.

\subsubsection{Beyond geometric optics}

At the first bGO order, by plugging Eq.~\eqref{eq:tetrad_decomposition_amplitude} into Eq.~\eqref{eq:transverse gauge bgo}, one gets
\begin{equation}
2\tilde{\alpha}^{(1)}_{nn}n_\mu + \tilde{\alpha}^{(1)}_{nl}l_\mu + \tilde{\alpha}^{(1)}_{mn}m_\mu + \tilde{\alpha}^{(1)}_{nk}k_\mu = -2i \nabla^\nu \tilde{h}^{(0)}_{\mu\nu}.
\end{equation}
These components are no longer constrained to vanish. However, imposing $n^\mu \tilde{h}^{(1)}_{\mu\nu} = 0$ and $\tilde{h}^{(1)} = 0$ removes
\begin{equation}
\tilde{\alpha}^{(1)}_{kk} = \tilde{\alpha}^{(1)}_{km} = \tilde{\alpha}^{(1)}_{kl} = \tilde{\alpha}^{(1)}_{kn} = \tilde{\alpha}^{(1)}_{ml} = 0,
\end{equation}
leaving
\begin{equation}\label{unconstrained modes}
\begin{split}
\tilde{h}^{(1)}_{\mu\nu} &= \tilde{\alpha}^{(1)}_{mm} m_\mu m_\nu + \tilde{\alpha}^{(1)}_{ll} l_\mu l_\nu + \tilde{\alpha}^{(1)}_{nn} n_\mu n_\nu \\
&\quad + \tilde{\alpha}^{(1)}_{nm} n_{(\mu} m_{\nu)} + \tilde{\alpha}^{(1)}_{nl} n_{(\mu} l_{\nu)}.
\end{split}
\end{equation}
These extra degrees of freedom represent new unconstrained polarization structures induced by subleading corrections.

Building on these considerations, one can develop a rather intuitive understanding of how corrections bGO corrections enter and affect the metric perturbation. Assuming that higher-order modes $\tilde{\alpha}_{AB}^{(1)}(\xi)$ in Eq.~\eqref{unconstrained modes} have been evaluated\footnote{For the explicit computation of such modes in symmetron gravity, see Sec.~\ref{homogeneous spherical lens}. For analogous results in GR and Brans–Dicke theory, see  \cite{Dalang:2021qhu} and \cite{Menadeo:2024uoq}, respectively.}, one can explicitly show, by recalling Eq.~\eqref{WKB decomposition metric} projected along the basis direction, that such corrections manifest as a phase shift in the propagating signal, namely

\begin{align}
    \hat{e}^\mu_A \hat{e}^\nu_B \tilde{h}_{\mu\nu} 
    &= \tilde{\alpha}^{(0)}_{AB} \left(1 + i \epsilon\, \frac{\tilde{\alpha}^{(1)}_{AB}}{\tilde{\alpha}^{(0)}_{AB}} + \cdots \right) e^{i \theta / \epsilon} \\
    \label{beta}
    &\approx \tilde{\alpha}^{(0)}_{AB} e^{i \theta / \epsilon} \exp\left(i \epsilon\, \frac{\beta_{AB}}{G M_{L} f}\right),
\end{align}
where the terms in parenthesis have been resummed to an exponential assuming a small correction \cite{Menadeo:2024uoq}. The second line defines the \emph{lens-induced dispersion} (LID) parameter
\begin{equation}
\label{betabox}
\boxed{\beta_{AB} = G M_{L} f \frac{\tilde{\alpha}^{(1)}_{AB}}{\tilde{\alpha}^{(0)}_{AB}}}\,.
\end{equation}
quantifying the dispersion of the signal and isolating it from the dependence on the signal frequency $ f $ and the characteristic lens mass scale $M_{ L}$. Note that $\beta_{AB}$ is a matrix in field space and the exponential ansatz relies on the fact that higher-order corrections follow the same functional structure. 

This expression demonstrates how the sub-leading corrections introduce a frequency- and polarization-dependent phase shift. As clarified in \cite{Menadeo:2024uoq}, dispersion is a promising test for modified gravity that can be performed on any GW signal, without the need for an electromagnetic counterpart.
The LID parameter is also a particular case of a parameterized post-Einsteinian deviation~\cite{Yunes:2009ke,Cornish:2011ys}, where the value is distributed according to the probability of encountering a lens able to produce it (e.g.~environmental or stochastic).

\section{GW propagation in symmetron gravity}\label{symmetron}
In this section, we review symmetron gravity and study GW propagation in the GO limit and beyond, before analyzing lens-induced dispersion in this setting.

\begin{figure}[t]
    \centering
    \begin{tikzpicture}
    \shade[inner color=blue!80, outer color=white, opacity=0.9] (0,0) circle (2);

    \path[fill=cyan, opacity=0.3] 
        (0,0) circle (2.5)
        (0,0) circle (2.1);

    \draw[red, line width=1.5pt, opacity=0.5] (0,0) circle (2.5);

    \draw[very thick] (-3,1.5) 
        .. controls (-2.2,1.5) and (-2,0.3) .. (-2,0)
        -- (2,0)
        .. controls (2,0.3) and (2.2,1.5) .. (3,1.5);

    \node at (0,-.5) {screened};
    \node at (0,-2.05) {unscreened};
    \node at (3.2,1.8) {$\phi_0$};
    \node at (0,2.1) {\footnotesize thin shell};
    \end{tikzpicture}

    \caption{Schematic representation of the scalar field profile $\phi$ in a partially screened spherical object. The screened interior is represented with a blue gradient. A thin shell (cyan) surrounds the core and marks the transition region, beyond which the field grows rapidly. For more details on this figure and a comparison with objects that are fully screened or unscreened, see \cite{Sakstein:2014jrq}.}
    \label{fig:schematic_phi_profile}
\end{figure}
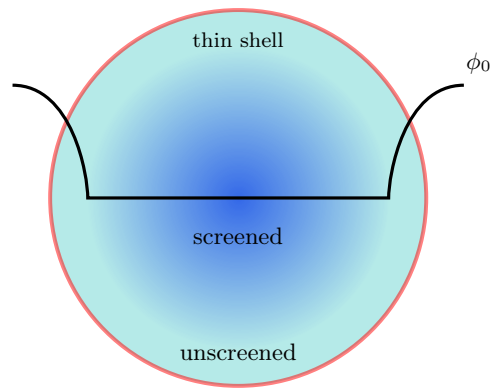

\subsection{Review of symmetron gravity}
The symmetron is a specific model of scalar-tensor theory that implements a density-dependent screening mechanism, namely the suppression of deviations from GR in highly dense environments. Unlike the chameleon, which screens through an environment-dependent effective mass, or the Vainshtein screening mechanism, which relies on the presence of non-linear derivative self-interactions of the scalar, the symmetron operates via spontaneous symmetry breaking. More specifically, there is a $\mathbb{Z}_2$ symmetry relating $\phi\rightarrow -\phi$ that is broken at low densities and restored at high densities. symmetron screening suppresses the matter coupling in high-density regions. 

In low-density regions, the symmetry is broken, the scalar field acquires a non-zero vacuum expectation value (VEV) and mediates a fifth force interaction beyond GR. In high-density environments, the symmetry is restored, the VEV vanishes, and the scalar interaction is suppressed.

In the next paragraph, we present the symmetron action in the Einstein frame, as in the original references \cite{Hinterbichler:2010es, Hinterbichler:2011ca}. Subsequently, we switch to the Jordan frame description and explain why, for the purposes of the present work, it is equivalent to perform the analysis in either frame. We will choose to work in the Jordan frame to make the comparison with the dispersive effects in Brans-Dicke \cite{Menadeo:2024uoq} theory easier.

The symmetron action in the Einstein frame reads \footnote{Unlike  \cite{Hinterbichler:2010es,Hinterbichler:2011ca,Burrage:2017qrf}, where a tilde is used to indicate Jordan frame quantities, we reserve the tilde notation for variables related to the trace-reversed metric perturbation, consistently with the convention adopted in  \cite{Menadeo:2024uoq}. In the present work, we use the subscript $^{\rm E}$ to indicate Einstein frame quantities and use standard notation for Jordan frame quantities.}
\begin{equation}
\label{eq:symmetron_action}
S^{\rm E}=\int \mathrm{d}^{4} x \sqrt{-g^{\rm E}}\left[\frac{M_{\rm P}^{2}}{2} \mathcal{R}^{\rm E}-\frac{1}{2}(\nabla \phi)^{2}-V^{\rm E}(\phi)\right]+S_{\mathrm{m}}[g],
\end{equation}

where $g^{\rm E}$ is the determinant of the Einstein metric $g_{\mu\nu}^{\rm E}$, $\mathcal{R}^{\rm E}$ is the Ricci scalar constructed from this metric, and $M_{\rm P}\equiv 1/\sqrt{8\pi G}$ is the reduced Planck mass, with $G$ the Newton constant. 
$V^{\rm E}(\phi)$ is a symmetry-breaking potential, the simplest example being
\begin{equation}
\label{pot}
    V^{\rm E}(\phi)=-\frac{\mu^{2} \phi^{2}}{2}+\frac{\lambda \phi^{4}}{4},
\end{equation}
where $\mu$ is the mass scale triggering the spontaneous symmetry breaking and $\lambda$ is a dimensionless self-coupling constant.
The matter action $S_{\mathrm{m}}$ is coupled to the Jordan frame metric $g_{\mu \nu}$, related to the Einstein frame metric $g_{\mu \nu}^{\rm E}$ by the conformal transformation 
\begin{equation}\label{conformal metric}
g_{\mu \nu}=A^{2}(\phi) g_{\mu \nu}^{\rm E}, 
\end{equation}
where $A(\phi)$ is the conformal coupling
\begin{equation}
\label{confo}
A(\phi) = 1 + \frac{\phi^2}{2M_s^2},
\end{equation}
and $M_s$ is the coupling between the symmetron field and matter.

The metric and scalar field equations are

\begin{equation}
M_{\rm P}^2 \mathcal{G}^{\rm E}_{\mu\nu} = T^{\rm E}_{\mu\nu} + A^2(\phi) T_{\mu\nu} \,,
\end{equation}
\begin{equation}\label{scalar_field_EOM_EF}
\square^{\rm E} \phi-V^{\rm E}_{, \phi}+A^{3}(\phi) A_{, \phi} T=0, 
\end{equation}
with 
\begin{equation}
T^{\rm E}_{\mu\nu} = \partial_\mu \phi \, \partial_\nu \phi - \frac{1}{2} g_{\mu\nu}^{\rm E} (\partial \phi)^2 - g_{\mu\nu}^{\rm E} V^{\rm E}(\phi) \,.
\end{equation}
Here, $\mathcal{G}^{\rm E}_{\mu\nu}\equiv \mathcal{R}^{\rm E}_{\mu\nu}-1/2 g^{\rm E}_{\mu\nu}\mathcal{R}^{\rm E}$ is the Einstein tensor and $T=T_{\mu \nu} g^{\mu \nu}$ is the trace of the energy-momentum tensor in the Jordan frame, 
which is covariantly conserved: ${\nabla}^{\mu} \tensor{{T}}{_\mu_\nu}=0$.

The independent parameters characterizing symmetron gravity are $\mu,\lambda$ and $M_s$. For local tests of gravity, non-linearities and scalar backreaction are ignored, and astrophysical objects can be modeled as spherically symmetric pressureless bodies. Therefore, Eq.~\eqref{scalar_field_EOM_EF} in spherical coordinates reads
\begin{equation}
\label{radial}
\frac{\mathrm{d}^{2}}{\mathrm{~d} r^{2}} \phi+\frac{2}{r} \frac{\mathrm{~d}}{\mathrm{~d} r} \phi=V^{\rm E}_{, \phi}+\rho^{\rm E} A_{, \phi} ,
\end{equation}
where we used the matter density $\rho^{\rm E}=A^{3} \rho$ conserved in the Einstein frame (we will see in the following that the distinction between the two becomes irrelevant for our purposes).

In the relevant case of the interior or exterior of a star or galaxy, $\rho^{\rm E}$ is fairly homogeneous and the field evolves following an effective potential
\begin{equation}
V^{\rm E}_{\mathrm{eff}}(\phi)=V^{\rm E}(\phi)+\rho^{\rm E} A(\phi). 
\end{equation}
Using Eqs.~\eqref{pot} and \eqref{confo}, the latter becomes
\begin{equation}\label{effective potential}
V^{\rm E}_{\mathrm{eff}}(\phi)=\frac{1}{2}\left(\frac{\rho^{\rm E}}{M_s^{2}}-\mu^{2}\right) \phi^{2}+\frac{1}{4} \lambda \phi^{4}, 
\end{equation}
which illustrates how the symmetron screening mechanism works. 
On the one hand, in vacuum or in cosmic voids, where $\rho^{\rm E} \simeq 0$, the symmetry is spontaneously broken and the scalar acquires a VEV $|\phi|=\phi_0 \equiv \mu / \sqrt{\lambda}$. On the other hand, in regions of high density, where $\rho^{\rm E}>M_s^2 \mu^2$, the effective potential no longer breaks the symmetry, and the VEV vanishes. In the first case, the astrophysical object is screened, meaning that the fifth force is suppressed, whereas in the second case, the object is unscreened and the scalar field mediates a non vanishing long-range interaction. In particular, the coupling is proportional to the local VEV and turns on in low-density environments, enabling the symmetron to couple to matter.
It is usually assumed that the symmetry breaks around the onset of cosmic acceleration \cite{Hinterbichler:2011ca} (note that the symmetron energy scale is too small for it to act as dark energy, thus a cosmological constant is still needed)
\begin{equation}\label{eq:mu from Ms}
H_{0}^{2} M_{\rm P}^{2} \sim \mu^{2} M_s^{2} \to \mu\sim H_0\frac{M_{\rm P}}{M_s},
\end{equation}
where $H_0$ is the Hubble constant, thereby reducing the number of independent parameters in symmetron gravity from three to two.

Local gravity tests require $M_s \lesssim$ $10^{-3} M_{\rm P}$, thus the range of the symmetron-mediated force in vacuum is $\lesssim$ Mpc. By employing Eq.~\eqref{eq:mu from Ms}, the scalar interaction is comparable to gravity when 
\begin{equation}
\phi_{0} \equiv \mu / \sqrt{\lambda}\sim \dfrac{M_s^{2}}{M_{\rm P}}.
\end{equation}
Further, the requirement of $M_s=10^{-3} M_{\rm P}$ together with Eq.~\eqref{eq:mu from Ms} fixes an upper limit for $\lambda$, namely
\begin{equation}\label{lambda max}
    \lambda\gtrsim 10^{-96} \,.
\end{equation}

A crucial feature of screening in symmetron solutions, similar to chameleon behavior \cite{Khoury:2003aq,Khoury:2003rn} is \textit{thin-shell screening}. Sufficiently massive ($\rho^{\rm E}>\mu^2M_s^2$) spherically symmetric sources exhibit weak coupling of the symmetron field to matter inside the core of the object, thus $\phi\approx 0$. Near the surface, the field profile grows substantially (see Figure \ref{fig:schematic_phi_profile}), connecting to the asymptotic behavior away from the object. The scalar field derivatives exhibit similar behavior when approaching the object's shell. This aspect will play a relevant role in Sec.~\ref{dispersive corrections in symmetron}. The exterior profile, instead, is dominated by the contribution near the surface: thin-shell screening is responsible for the suppression of the symmetron effects outside the object. The strength of screening is quantified by the parameter
\begin{equation}
\label{alpha}
    \alpha\equiv\dfrac{\rho^{\rm E} R^2}{M_s^2}=6 \frac{M_{\rm P}^2}{M_s^2} \Phi_{\rm N},
\end{equation}
where $R$ is the radius of the object and $\Phi_{\rm N}$ is the Newtonian potential at its surface. The screening parameter $\alpha$
is the inverse of the so-called thin-shell factor $\Delta R/R\equiv M_s^2/\rho^{\rm E} R^2$. Sufficiently massive objects with $\alpha\gg 1$ exhibit the thin-shell effect and are screened, while smaller objects with $\alpha\ll 1$ do not (and in this case the symmetron-mediated scalar force gives $\mathcal{O}(1)$ contributions to the gravitational potential).

\subsubsection{Scalar field solutions}\label{sec:Scalar field solutions}

\begin{figure*}[t]
    {
    \includegraphics[width=\textwidth]{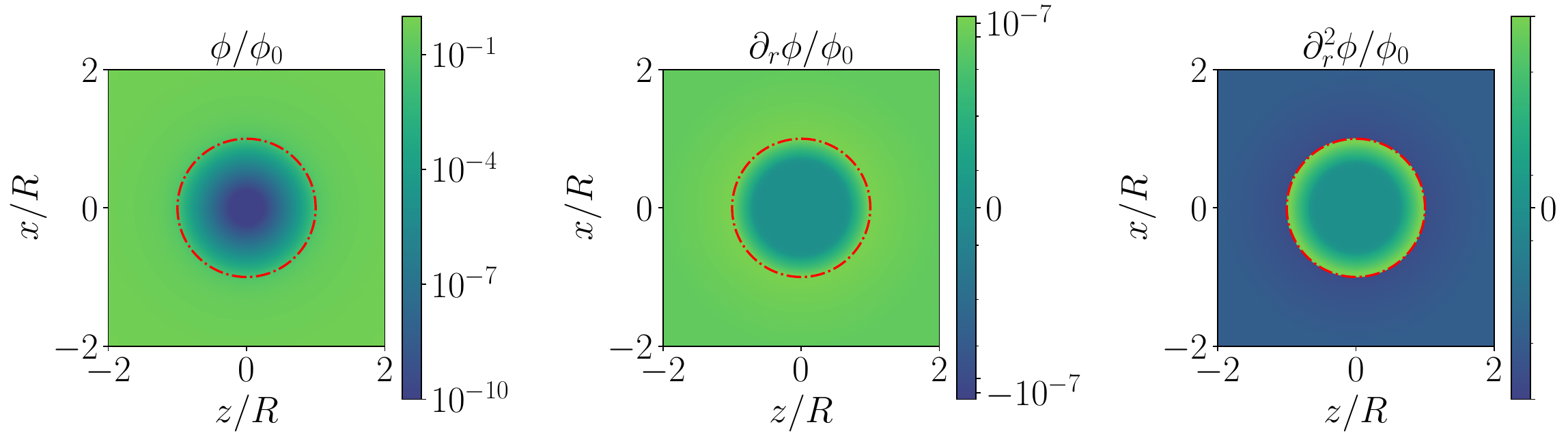}}
    \caption{Normalized scalar field profile $\phi/\phi_0$ (left panel) and its radial derivatives (central and right panel) for a spherical Earth-like object. The red dashed circle indicates the boundary $r = R$. The values $\lambda = 10^{-80}$ and $M_s = 10^{12}~\mathrm{GeV}$ are adopted as illustrative benchmarks to show the spatial structure of the field and its gradients. As $M_s$ increases and $\lambda$ pushed towards its limit value bound (see Eq.~\eqref{lambda max}), the smooth thin-shell region visible in the left panel becomes increasingly narrow. This implies a more abrupt variation of the scalar field profile and its derivatives at the boundary.}
    \label{fig:scalar_field}
\end{figure*}

\twocolumngrid 

The solutions of Eq.~\eqref{radial} are provided in \cite{Hinterbichler:2010es,Hinterbichler:2011ca}. Being a second-order differential equation, the equation requires two boundary conditions to admit a unique solution: regularity at the center $\partial_r\phi_{(r=0)} = 0$, and asymptotic matching to the vacuum expectation value $\phi_{(r \to \infty)} = \phi_0$.

\textbf{Inside solution.} In the region $r < R$, where $\rho^{\rm E} \gg \mu^2 M_s^2$, the effective potential~\eqref{effective potential} can be approximated as quadratic around its minimum, $V^{\rm E}_{\rm eff}\simeq \rho^{\rm E} \phi^2/2M_s^2$. The scalar field profile inside the object is then given by
\begin{equation}\label{phi in}
\phi_{\mathrm{in}}(r)=A \frac{R}{r} \sinh \left(r \sqrt{\frac{\rho^{\rm E}}{M_s^{2}}-\mu^{2}}\right),
\end{equation}
where $A$ is an arbitrary constant, determined by matching the interior to the exterior solution (as discussed below).

\textbf{Outside solution.} In the region $r > R$, where the matter density drops below the critical threshold, the effective potential can be approximated as quadratic around the symmetry-breaking VEV $\phi_0$, thus reading $V^{\rm E}_{\rm eff}(\phi) \simeq \mu^2 (\phi - \phi_0)^2/2$. Under this assumption, the scalar field profile outside the object becomes
\begin{equation}\label{phi out}
\phi_{\mathrm{out}}(r)=B \frac{R}{r} e^{-\sqrt{2} \mu (r-R)}+\phi_{0},
\end{equation}
with $B$ an undetermined constant.  

Requiring continuity of the scalar field and its first derivative at the boundary $r = R$ uniquely fixes the integration constants $A$ and $B$. These coefficients can be compactly expressed in terms of the screening parameter $\alpha$ introduced in Eq.~\eqref{alpha}, leading to
\begin{align}
A &= \phi_0 \sqrt{\frac{1}{\alpha}}\, \text{sech}\left(\sqrt{\alpha}\right) \,, \\
B &= -\phi_0 \left[1 - \sqrt{\frac{1}{\alpha}}\, \tanh\left(\sqrt{\alpha}\right)\right] \,.
\end{align}
The behavior of the coefficients for $\alpha \gg 1$ (strong screening) and $\alpha \ll 1$ (no screening) can be found in  \cite{Hinterbichler:2010es}.

\subsubsection{Jordan frame description}
In the Einstein frame action \eqref{eq:symmetron_action}, matter moves on geodesics of the Jordan frame. Switching to the Jordan frame through the conformal transformation given by Eq.~\eqref{conformal metric}, matter still moves on Jordan frame geodesics, but gravitational potentials are modified through mixing with the scalar. Since our aim in this work is to connect with physical observables, we choose to work in the Jordan frame. In this frame, the symmetron field is non-minimally coupled to gravity and minimally coupled to matter, and essentially behaves like a generalized Brans-Dicke theory with a potential and additional matter fields. Working in the Jordan frame also ensures a straightforward comparison of our results with those for Brans-Dicke theory found in \cite{Menadeo:2024uoq}.

The symmetron action, in terms of the Jordan-frame metric~\eqref{conformal metric} is \footnote{Switching from the Einstein to the Jordan frame also entails a field redefinition, where $\varphi\equiv A^{-2}(\phi^{E})$ is the Jordan frame scalar field and $\phi^{\rm E}$ the Einstein frame one, which has been hitherto denoted with $\phi$. However, since we will restrict to the Newtonian limit, as explained in the next paragraph, $A(\phi)$ is usually taken to be unity for all relevant field values \cite{Hinterbichler:2011ca}, and in the following we only use $\phi$ for simplicity.} \cite{Burrage:2017qrf}
\begin{equation}\label{eq:symmetron_actionJF}
\begin{split}
 S &=\int d^4x \sqrt{-g} \left[\dfrac{M_{\rm P}^2}{2A^2(\phi)} \mathcal{R} - \dfrac{k(\phi)}{2}\, (\nabla\phi)^2 - V(\phi) \right] +\\&+S_{\rm m}[g^{\rm E}],   
\end{split}
\end{equation}
where $V(\phi)\equiv V^{\rm E}(\phi)/{A^4(\phi)}$ and the non-canonical kinetic coefficient $k(\phi)$ is defined as

\begin{equation}
k(\phi) = \frac{1}{A^2(\phi)} \left[ 1 - 6 M^2_{\rm P}\left( \frac{d \ln A(\phi)}{d\phi} \right)^2 \right].
\end{equation}

In the Jordan frame formulation, the conformal coupling function $A^2(\phi)$, introduced in Eq.~\eqref{confo}, governs the non-minimal interaction between the scalar field and the Ricci scalar in the gravitational Lagrangian. By construction, $A^2(\phi) \to 1$ for large $M_s$ values (or, equivalently, in the limit $M_s \to \infty$), effectively suppressing the coupling and recovering a minimally coupled theory, as in GR (provided the scalar field becomes non-dynamical or vanishes). Conversely, for smaller values of $M_s$, the function $A^2(\phi)$ deviates more significantly from unity, enhancing the coupling between the scalar field and spacetime curvature.

The metric and scalar field equations in this frame read
\begin{widetext}
\begin{equation}\label{eq:metric_field_equation}
{\mathcal{G}}_{\mu\nu} = \frac{A^2(\phi)}{M_{\rm P}^2} \left[
T^{\rm m}_{\mu \nu}
+ k(\phi)\, \nabla_\mu \phi\, \nabla_\nu \phi
- g_{\mu \nu} \left( \frac{k(\phi)}{2} \nabla_\alpha \phi\, \nabla^\alpha \phi + V(\phi) \right)
+ \left( \nabla_\mu \nabla_\nu - g_{\mu \nu} \Box \right) A^{-2}(\phi)
\right]
\end{equation}
\begin{equation}\label{eq:scalar_field_equation}
k(\phi) \Box \phi - V_{\phi} + k_{,\phi} \nabla^\alpha \phi\, \nabla_\alpha \phi -\dfrac{1}{2} M^2_{\rm P}\frac{A_{,\phi}}{A^3} \mathcal{R} = 0,
\end{equation}
\end{widetext}
where $T^{\rm m}_{\mu\nu}$ is the stress-energy tensor related to the matter field and the subscript $_{,\phi}\equiv\partial_{\phi}$
.
\subsubsection{Equivalence of frames in Newtonian limit}
The dynamics of the scalar field in the Jordan frame is different than in the Einstein frame. However, since we are interested in studying a static matter distribution in the universe, far away from the source emitting the gravitational waves, we are justified in taking the Newtonian limit, which greatly simplifies the expressions above. In the following, we show that in this limit, the scalar field equation in the Jordan frame and in the Einstein frame takes the same form \cite{Burrage:2017qrf}. Non-relativistic screening, therefore, works identically in both frames.
Specifically, in this limit, we neglect terms at orders higher than $v^2 / c^2$, such as $ (\partial \phi)^2 \sim \mathcal{O}\left(\dfrac{v^4}{c^4}\right)$, and time derivatives of the scalar field.
We can approximate
\begin{equation}
V^{\rm E}(\phi) \approx V(\phi), \quad T^{\mathrm{m}} \approx-\rho,
\end{equation}
and find 
\begin{equation}
    \mathcal{R} \approx\dfrac{-T^{\mathrm{m}}A(\phi)^2}{M_{\rm P}^2} \approx \dfrac{\rho A(\phi)^2}{M_{\rm P}^2}.
\end{equation}
Since $T_{\mathrm{m}}^{\mu \nu}=A^{-6} T_{\mathrm{m}}^{\mu \nu\, \rm E}$, one has $T_{\mathrm{m}}=A^{-4} T^{\rm E}_{\mathrm{m}}$, so that $\rho=\rho^{\rm E}+\mathcal{O}\left(v^4 / c^4\right)$. With these approximations, the scalar field equation in the Jordan frame (in radial coordinates) \eqref{eq:scalar_field_equation} is exactly the same as in the Einstein frame \eqref{radial} 
\begin{equation}
\frac{\mathrm{d}^{2}}{\mathrm{~d} r^{2}} \phi+\frac{2}{r} \frac{\mathrm{~d}}{\mathrm{~d} r} \phi=V_{, \phi}+A_{, \phi} \rho.
\end{equation}
This means that spherically symmetric solutions inside \eqref{phi in} and outside \eqref{phi out} a spherical object can be used in both Einstein and Jordan frames.
We thus perform all the following calculations in the Jordan frame.

\subsection{Short-wave expansion for symmetron gravity}
The GW propagation equation in symmetron gravity can be derived by perturbing the field equations~\eqref{eq:metric_field_equation} at first order, using Eq.~\eqref{metric perturbation}. The resulting wave equation, written in terms of the trace-reversed metric perturbation $\tilde{h}_{\mu\nu}$, takes the form given in Eq.~\eqref{eq:propagation_equation}. The expressions for $\tensor{\mathsf{K}}{_\mu_\nu^\alpha^\beta^\gamma^\rho}$, $\tensor{\mathsf{A}}{_{\mu\nu}^{\alpha\beta\gamma}}$, and $\tensor{\mathsf{M}}{_{\mu\nu}^{\alpha\beta}}$, specific for symmetron gravity, are provided in Appendix~\ref{appendix:symmetron_functions}. As already mentioned at the beginning of Sec.~\ref{General formalism for GW propagation}, we do not perturb the scalar field equation~\eqref{eq:scalar_field_equation}, since scalar waves are not considered in this work.

Let us now apply the formalism discussed in Section~\ref{shortwave} to the symmetron case.
\subsubsection{Geometric optics}
To determine the evolution of the tensor amplitude components, we insert Eq.~\eqref{eq:tetrad_decomposition_amplitude} into the generalized transport equation~\eqref{eq:eps1} and project onto $\hat{e}^\mu_A \hat{e}^\nu_B$, obtaining
\begin{equation}
\mathcal{D} \left( \frac{\tilde{\alpha}^{(0)}_{mm}(\xi)}{A[\phi(\xi)]} \right) = \mathcal{D} \left( \frac{\tilde{\alpha}^{(0)}_{ll}(\xi)}{A[\phi(\xi)]} \right) = 0,
\end{equation}
and the differential operator is defined as
\begin{equation}
\mathcal{D} \equiv 2k^\alpha \nabla_\alpha + \nabla_\alpha k^\alpha.
\end{equation}
Since $k^\mu$ is tangent to the null geodesic parametrized by the affine parameter $\xi$, the derivative along the ray is given by $k^\mu \nabla_\mu = d/d\xi$. Furthermore, the divergence of the null vector can be related to the comoving  distance along the geodesic $D(\xi)$ via the identity $\nabla_\mu k^\mu = 2 \, d\ln D(\xi) / d\xi$, which follows from the Raychaudhuri equation in absence of shear and rotation (see discussion in \cite{Menadeo:2024uoq}). Therefore, the transport operator reduces to a total derivative
\begin{equation}
\frac{d}{d\xi} \left[ \frac{\tilde{\alpha}^{(0)}_{\circ} D(\xi)}{A[\phi(\xi)]} \right] = 0,
\end{equation}
where $\circ \in \{mm, ll\}$. This integrates to
\begin{equation}\label{GO amplitude evolution symmetron}
\tilde{\alpha}^{(0)}_{\circ}(\xi) = \left( \frac{A[\phi(\xi)]}{D(\xi)} \right) \frac{D(\xi_s)}{A[\phi(\xi_s)]} \, \tilde{\alpha}^{(0)}_{\circ}(\xi_s),
\end{equation}
and the full amplitude evolves as
\begin{equation}
\begin{split}
\tilde{h}^{(0)}_{\mu\nu}(\xi) &= \left( \frac{A[\phi(\xi)]}{D(\xi)} \right) \frac{D(\xi_s)}{A[\phi(\xi_s)]} \\
&\quad \times \left( \tilde{\alpha}^{(0)}_{mm}(\xi_s) m_\mu m_\nu + \tilde{\alpha}^{(0)}_{ll}(\xi_s) l_\mu l_\nu \right).
\end{split}
\end{equation}

\subsubsection{Beyond geometric optics}

Their evolution is obtained by integrating Eq.~\eqref{eq:eps0} along the parameter $\xi$, which parametrizes the signal's geodesic. This is done using the decomposition in Eq.~\eqref{eq:tetrad_decomposition_amplitude}, followed by a projection onto the null tetrad (dual) basis $\hat{e}^\mu_A \hat{e}^\nu_B$. The result is
\begin{equation}\label{bGO formal symmetron solution}
\begin{split}
&\tilde{\alpha}_{AB}^{(1)}(\xi) = \left( \frac{A[\phi(\xi)]}{D(\xi)} \right) \frac{D(\xi_s)}{A[\phi(\xi_s)]} \, \tilde{\alpha}^{(1)}_{AB}(\xi_s)+ \\
&\quad + i \frac{A[\phi(\xi)]}{D(\xi)} \int_{\xi_s}^{\xi} d\xi' \, \hat{e}^\mu_A \hat{e}^\nu_B A[\phi(\xi')] D(\xi')\left(  \mathsf{P}_{\mu\nu}^{\alpha\beta} \tilde{h}^{(0)}_{\alpha\beta}\right).
\end{split}
\end{equation}
The imaginary unit appearing in front of the second term in the equation above indicates that this contribution affects the phase of the signal. This is sourced by the propagation operator~\eqref{eq:propagator_operator} acting on the geometric optics amplitude. These phase modifications are intrinsically frequency-dependent due to the nature of the short-wavelength expansion introduced in Eq.~\eqref{WKB decomposition metric}.

\section{Dispersion induced by a homogenous spherical lens}\label{homogeneous spherical lens}
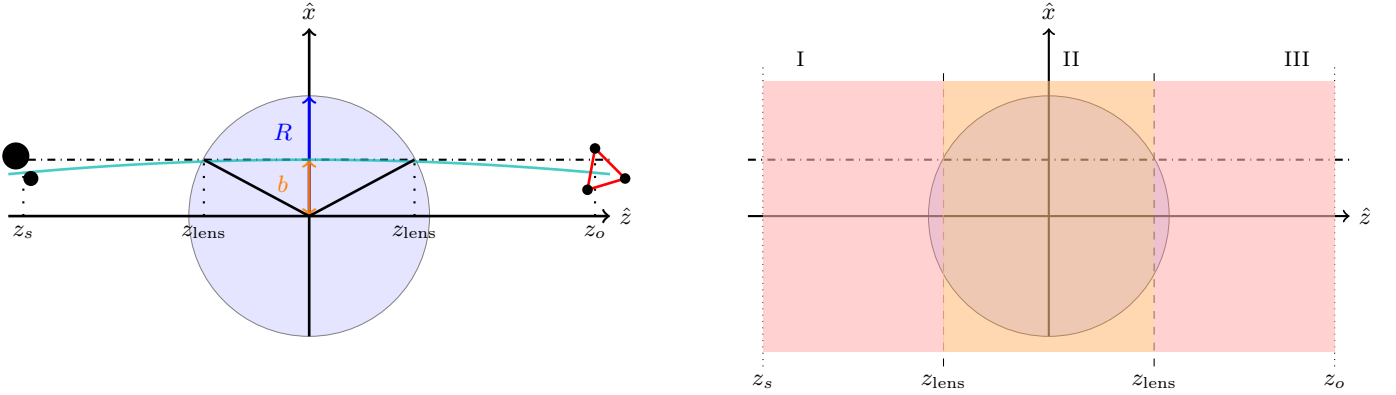
\begin{figure*}[]
    \centering

    \begin{subfigure}[t]{.45\textwidth}
        \centering
        \begin{tikzpicture}[baseline=(zaxis.base)]
            \node (zaxis) at (0,0) {};

            \filldraw[fill=white!80!blue, draw=black, line width=0pt, opacity=0.5] (0,0) circle (1.6);

            \draw[->, black, line width=1pt] (-4,0)--(4,0) node[right]{$\hat{z}$};
            \draw[->, black, line width=1pt] (0,-1.6)--(0,2.5) node[above]{$\hat{x}$};

            \draw[->, blue, line width=1pt] (0,0) -- (0,1.6);
            \node at (-.35,1.35) [below] {\color{blue} $R$};
            \draw[<->, orange, line width=1pt,opacity=0.9] (0,0) -- (0,.75);
            \node at (-.35,.65) [below] {\color{orange} $b$};

            \draw[loosely dotted, black, line width=0.8pt] (-3.8,0)--(-3.8,.75) (-3.8,0) node[below]{$z_s$};
            \draw[loosely dotted, black, line width=0.8pt] (3.8,0)--(3.8,.75) (3.8,0) node[below]{$z_o$};
            \draw[dashdotted, black, line width=0.8pt] (-4,0.75)--(4,.75);

            \draw[domain=-4:4, cyan,opacity=0.75, line width=1pt] plot (\x, {-0.012*(\x)^2+.75});

            \fill[black] (-3.9,.8) circle (.18);
            \fill[black] (-3.7,.5) circle (.1);

            \coordinate (A) at (3.7,.35);
            \coordinate (B) at (4.2,.5);
            \coordinate (C) at (3.8,.9);
            \draw[red, line width=1pt] (A) -- (B) -- (C) -- cycle;
            \fill[black] (A) circle (2pt);
            \fill[black] (B) circle (2pt);
            \fill[black] (C) circle (2pt);

            \draw[black, line width=1pt] (0,0) -- (1.4,.75) node[midway, above] {};
            \draw[black, line width=1pt] (0,0) -- (-1.4,.75) node[midway, above] {};
            \draw[loosely dotted, black, line width=0.8pt] (-1.4,0)--(-1.4,.75) (-1.4,0) node[below]{$z_{\text{lens}}$};
            \draw[loosely dotted, black, line width=0.8pt] (1.4,0)--(1.4,.75) (1.4,0) node[below]{$z_{\text{lens}}$};
        \end{tikzpicture}
        \label{fig:panel_a_propagation}
    \end{subfigure}
    \hfill
    \begin{subfigure}[t]{.45\textwidth}
        \centering
        \begin{tikzpicture}[baseline=(zaxis.base)]
            \node (zaxis) at (0,0) {};

            \filldraw[fill=white!80!blue, draw=black, line width=0pt, opacity=0.8] (0,0) circle (1.6);
            
            \draw[dashdotted, black, line width=0.8pt] (-4,0.75)--(4,.75);

            \draw[->, thick] (-4,0) -- (4,0) node[anchor=west] {$\hat{z}$};
            \draw[->, thick] (0,-1.6) -- (0,2.5) node[anchor=south] {$\hat{x}$};

            \draw[dashed] (-1.4,-2) -- (-1.4,2);
            \node at (-1.4,-2.2) {$z_{\text{lens}}$};

            \draw[dashed] (1.4,-2) -- (1.4,2);
            \node at (1.4,-2.2) {$z_{\text{lens}}$};

            \draw[dotted] (-3.8,-2) -- (-3.8,2);
            \node at (-3.8,-2.2) {$z_s$};

            \draw[dotted] (3.8,-2) -- (3.8,2);
            \node at (3.8,-2.2) {$z_o$};

            \fill[red!30, opacity=0.6] (-3.8,-1.8) rectangle (-1.4,1.8);
            \fill[orange!50, opacity=0.6] (-1.4,-1.8) rectangle (1.4,1.8);
            \fill[red!30, opacity=0.6] (1.4,-1.8) rectangle (3.8,1.8);

            \node at (-3.3,2.1) {\footnotesize I};
            \node at (0.3,2.1) {\footnotesize II};
            \node at (3.3,2.1) {\footnotesize III};
        \end{tikzpicture}
        \label{fig:panel_b_propagation_regions}
    \end{subfigure}

\caption{
    Schematic representation of the wave propagation setup. \textbf{Left panel.} The wave is emitted at $z_s$, propagates along the $\hat{z}$-direction with an impact parameter $b$, and crosses a spherically symmetric lens located at $z_{\text{lens}}$. The propagation path is split into regions inside and outside the lens (with $r$ the radial coordinate from the lens center). \textbf{Right panel.} The entire propagation path is divided into three regions: I, before encountering the lens; II, inside the lens (i.e., $|z - z_{\text{lens}}| < R$); and III, after encountering the lens. Each region is treated independently to evaluate the corresponding contribution to the total signal.
    }    \label{fig:propagation_panels}
\end{figure*}
In \cite{Menadeo:2024uoq}, the authors considered only a point-like lens and derived the dispersion effects it causes on GWs in Brans-Dicke theory. In this work, we go beyond this simplification and consider a more realistic spherically symmetric lens with constant density. The symmetron action \eqref{eq:symmetron_action} naturally incorporates the baryonic matter fields that make up the astrophysical object acting as the lens.
It is not \textit{a priori} expected that, only because the inside of the object is screened, symmetron effects are not present and automatically reduce to GR effects. The fact that the scalar field profile inside the object is not constant, allows its gradient to drive the dispersive effects. Additionally, in Section \ref{dispersive corrections in GR} we will show that similar (albeit weaker) dispersive effects are also present in GR with an extended lens. These dispersive effects show up concretely as a frequency-dependent phase shift in the waveform.

In the following, we evaluate the dispersive corrections on the GW signal induced by a spherically symmetric lens, whose line element can be described by a weak-field limit metric
\begin{equation}\label{wfl metric}
    ds^2 = -\left(1 + 2\Psi\right)dt^2 + \left(1 - 2\Psi\right)(dx^2 + dy^2 + dz^2),
\end{equation}
with $\Psi$ denoting the Newtonian gravitational potential for a lens of mass $M$ and radius $R$, defined as

\begin{numcases}{\Psi(r) = }
-\dfrac{G M}{2 R^3} \left(3 R^2 - r^2\right), & $r \leq R$\label{eq:grav_potential_in}\\
-\dfrac{G M}{r}, & $r > R$\label{eq:grav_potential_out},
\end{numcases}

where $ r = \sqrt{x^2 + y^2 + z^2} $ is the radial coordinate. For simplicity, and without loss of generality, one can fix $ y = 0$, so that the signal propagates entirely within the $(x,z)$-plane (see left panel of Figure \ref{fig:propagation_panels}). In this case, $x$ is the impact parameter, labeled as $x\equiv b$.

The procedure to compute the bGO correction is to perturbatively expand all the quantities involved in the calculation in Section \ref{General formalism for GW propagation}, at first order in the potential $\Psi$ (assuming $\Psi \ll 1$). This approach allows us to work linearly in the metric perturbation and leads to a first-order expression for the tensor amplitude \cite{Dalang:2021qhu}:
\begin{equation}\label{perturbed amplitudes WFL new}
\tilde{h}^{(n)}_{\mu\nu} = \bar{\tilde{\alpha}}^{(n)}_{AB} \bar{e}^A_{(\mu} \bar{e}^B_{\nu)} 
+ 2 \bar{\tilde{\alpha}}^{(n)}_{AB} \delta e^A_{(\mu} \bar{e}^B_{\nu)} 
+ \delta \tilde{\alpha}^{(n)}_{AB} \bar{e}^A_{(\mu} \bar{e}^B_{\nu)},
\end{equation}
where $\bar{\tilde{\alpha}}^{(n)}_{AB}$ and $\delta \tilde{\alpha}^{(n)}_{AB}$ represent, respectively, the unperturbed (background) and perturbed components of the amplitude in a tetrad basis, and the variation $\delta e^A_\mu$ captures the lens-induced modification of the tetrads (all background quantities are denoted with a bar). The unperturbed background tetrad can be taken as constant and parallel transported along the unperturbed trajectory, i.e. the $\hat{z}$-direction~(see  \cite{Will:2018bme} for more details)
\begin{align}
   \bar{k}^{\mu} &= \Omega (1, 0, 0, 1),           &\quad \bar{n}^{\mu} &= \frac{1}{2\Omega} (1, 0, 0, -1), \label{eq:tetrad_kn} \\
   \bar{m}^{\mu} &= \frac{1}{\sqrt{2}} (0, 1, i, 0), &\quad \bar{l}^{\mu} &= \frac{1}{\sqrt{2}} (0, 1, -i, 0). \label{eq:tetrad_ml}
\end{align}

Since one can reasonably assume that the deflection due to the lens is small, the signal’s trajectory deviates only slightly from the straight-line geodesic. Therefore, one can adopt the Born approximation and compute all corrections along the unperturbed path. Under this assumption, the  trajectory is parametrized with the affine parameter $d\xi = dz/\Omega$, where $\Omega = 2\pi/\lambda$ is the constant magnitude of the wavevector, and $\lambda$ is the wavelength of the signal.

The tetrad perturbation $\delta e^A_\mu$ at the first order in $\Psi$ is evaluated by integrating the linearized geodesic equation given by\til\cite{Cusin:2017fwz}, yielding
\begin{align}\label{eq: perturbed_tetrad_general}
\begin{split}
    \delta e^{\mu}_A&=-\int_{-\infty}^{+\infty
    }\frac{dz}{\Omega}\delta\Gamma^{\mu}_{\alpha\beta}\bar{e}^{\alpha}_A\bar{k}^{\beta}.
\end{split}
\end{align}

Since the perturbed tetrads are constructed from the linearized Christoffel symbols $\delta\Gamma^{\mu}_{\alpha\beta}$ in $\Psi$, their components are directly dependent on the gradient of the gravitational potential. Therefore, they are sensitive to the changes to $\Psi(r)$ encountered along the propagation path. In scenarios where the signal crosses the lens $b < R$ (see left panel of Figure \ref{fig:propagation_panels}), the geodesic traverses regions with different physical properties: the dense interior of the lens, characterized by a constant mass density and a quadratic gravitational potential~\eqref{eq:grav_potential_in}, and the surrounding vacuum, where $\Psi$ follows an inverse radial profile~\eqref{eq:grav_potential_out}. To properly account for these discontinuities in the gravitational potential, it becomes necessary to split the GW trajectory into three distinct propagation regions (see Figure \ref{fig:propagation_panels}):
\begin{itemize}
    \item \textbf{Region I}: from the source at \(z = -z_s\) to the entry point of the lens at \(z = -z_{\rm lens}\), where the signal propagates entirely outside the lens;
    \item \textbf{Region II}: from \(z = -z_{\rm lens}\) to \(z = +z_{\rm lens}\), where the signal crosses the interior of the lens and is affected by the internal gravitational potential~\eqref{eq:grav_potential_in};
    \item \textbf{Region III}: from \(z = +z_{\rm lens}\) to the observer at \(z = +z_o\), where the propagation occurs once again in vacuum.
\end{itemize}
\begin{figure*}[t]
    \centering
\includegraphics[width=\textwidth]{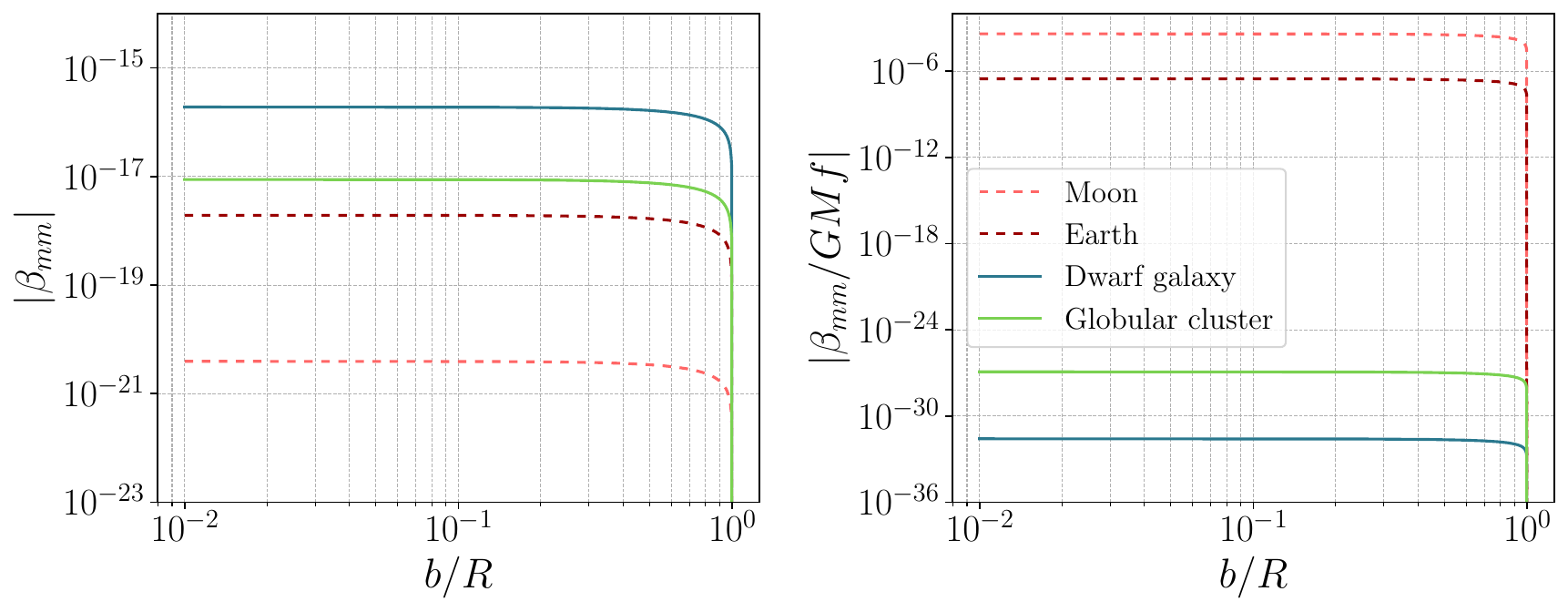}
    \caption{
    Evolution of the dispersion parameter $\beta_{mm}$, in GR, as a function of the normalized impact parameter $b/R$ for various astrophysical lenses. Dashed lines correspond to near-field objects, specifically the Earth and the Moon, which violate the short-wave condition $GMf \gtrsim 1$ (see Sec.~\ref{numerical results} for a detailed discussion on this). Solid lines, represent the results of more more massive objects for which $GMf \ll 1$ holds at $f = 100~\mathrm{Hz}$, when approximated as homogeneous spheres (see text).}
    \label{otherobjects}
\end{figure*}
In this setting, $z_{\rm lens}\equiv\sqrt{R^2-b^2}$, as it can be visually seen in the left panel of Fig.~\ref{fig:propagation_panels}. The perturbed tetrads then split into three respective contributions. For regions I and III, which lie entirely outside the matter distribution, the potential reduces to that of a point-like mass and the expressions follow from Eq.~\eqref{eq: perturbed_tetrad_general} with Eq.~\eqref{eq:grav_potential_out}:
\begin{align}
   \delta{k}^{\mu}_{\text{I/III}} &= \Omega \left(0, -\frac{4GM}{b}, 0, 0\right),  \\         
   \delta{n}^{\mu}_{\text{I/III}} &= (0, 0, 0, 0), \label{eq:tetrad_kn} \\
   \delta{m}^{\mu}_{\text{I/III}} &=\delta{l}^{\mu}_{\text{I/III}}= \frac{4GM}{\sqrt{2}} \left(-\frac{1}{b}, 0, 0, \frac{1}{b}\right), \label{eq:tetrad_ml}
\end{align}

where $R_s\equiv2GM$ is the Schwarzchild radius of the lens. Inside the lens (Region II), the potential changes and the expressions become
\begin{align}
   \delta{k}^{\mu}_{\text{II}} &= \Omega \left(0, -\frac{4 GM z_{\text{lens}}b}{R^3}, 0, 0\right),  \\         
   \delta{n}^{\mu}_{\text{II}} &= (0, 0, 0, 0), \\
   \delta{m}^{\mu}_{\text{II}} &=\delta{l}^{\mu}_{\text{II}}= \frac{4 GM}{\sqrt{2}} \left(-\frac{2z_{\text{lens}}b}{R^3}, 0, 0, \frac{2z_{\text{lens}}b}{R^3}\right). \label{eq:tetrad_ml_II}
\end{align}

As the impact parameter approaches the lens radius (\(b \to R\)), the segment of the path crossing the interior of the lens shrinks, and the entry and exit points \(z = \pm z_{\rm lens}\) tend to each other. In the strict limit \(b = R\), Region II vanishes entirely and the entire trajectory lies in vacuum, effectively reducing the setup to a point-lens configuration. In this limit, the configuration matches the point-lens scenario previously studied in GR and Brans-Dicke gravity, reported in  \cite{Cusin:2019rmt,Dalang:2021qhu} and  \cite{Menadeo:2024uoq}, respectively.

\subsection{General Relativity}\label{dispersive corrections in GR}

Before discussing the bGO corrections induced by a spherically symmetric lens in symmetron gravity, let us evaluate LID in GR considering as lenses the Earth and the Moon. The propagation equation for GWs in GR, using the TT gauge conditions~\eqref{transverse condition} and~\eqref{traceless condition}, reads
\begin{equation}
\Box\tilde{h}_{\mu\nu} - 2\tensor{\mathcal{R}}{^\alpha_\mu_\nu^\beta} \tilde{h}_{\alpha\beta} = 0.
\end{equation}

Inserting Eq.~\eqref{WKB decomposition metric} into the above wave equation and collecting terms order by order in the $\epsilon$ parameter, one obtains a hierarchy of equations for the phase and amplitude of the wave. Specifically, the first three orders correspond to
\\
\begin{itemize}
    \item $\mathcal{O}(\epsilon^{-2})$: the eikonal equation, which yields the null condition $k^\mu k_\mu = 0$;
    \item $\mathcal{O}(\epsilon^{-1})$: the transport equation for the leading-order GO amplitude;
    \item $\mathcal{O}(\epsilon^{0})$: the equation for the first-order bGO correction.
\end{itemize}

In particular, following the procedure of Sec.~\ref{General formalism for GW propagation}, the evolution of the GO amplitude in GR is governed by
\begin{equation}
\mathcal{D}\left(\tilde{\alpha}^{(0)}_{mm}\right) = \mathcal{D}\left(\tilde{\alpha}^{(0)}_{ll}\right) = 0 \quad \to \quad \frac{d}{d\xi}\left[\tilde{\alpha}^{(0)}_{\circ}D\right]=0,
\end{equation}
whose solution is
\begin{equation}
\tilde{\alpha}^{(0)}_{\circ}(\xi) = \frac{D(\xi_s)}{D(\xi)} \tilde{\alpha}^{(0)}_{\circ}(\xi_s).
\end{equation}
In GR, the amplitude evolution is only modulated by the distance along the background geodesic. In contrast, the equation for the symmetron case, \eqref{GO amplitude evolution symmetron}, shows that the amplitude is also influenced by the presence of the scalar field. This feature is well-known in the literature and has been already discussed in  \cite{Menadeo:2024uoq,Dalang:2021qhu,Dalang:2020eaj,Garoffolo:2019mna}.
At next-to-leading order, $\mathcal{O}(\epsilon^0)$, the equation for the bGO correction is
\begin{equation}
\mathcal{D}\tilde{h}^{(1)}_{\mu\nu} = i\left( \Box\tilde{h}^{(0)}_{\mu\nu} - 2\mathcal{R}_{\alpha\mu\nu\beta} \tilde{h}^{(0)\alpha\beta} \right),
\end{equation}
whose formal solution for the polarization amplitudes is given by:
\begin{equation}\label{BGO GR integration}
\begin{split}
\tilde{\alpha}_{AB}^{(1)}(\xi) &= \frac{D(\xi_s)}{D(\xi)} \tilde{\alpha}^{(1)}_{AB}(\xi_s)+ \\
&\quad + \frac{i}{D(\xi)} \int_{\xi_s}^{\xi} d\xi\, \hat{e}^{\mu}_{A} \hat{e}^{\nu}_{B} D(\xi) \left[ \Box\tilde{h}^{(0)}_{\mu\nu} - 2\mathcal{R}_{\alpha\mu\nu\beta} \tilde{h}^{(0)\alpha\beta} \right].
\end{split}
\end{equation}

Perturbing Eq.~\eqref{BGO GR integration} at first order in the gravitational potential $\Psi$ and translating the integration path along the $\hat{z}$-axis via the Born approximation, we obtain 
\begin{widetext}
\begin{equation}
\delta \tilde{\alpha}^{(1)}_{AB}(z_o) = \frac{i}{\bar{D}(z_o)} \left( \int_{-\infty}^{-z_{\rm lens}} + \int_{-z_{\rm lens}}^{+z_{\rm lens}} + \int_{+z_{\rm lens}}^{+\infty} \right)
\frac{dz}{\Omega}\, \hat{e}^\alpha_A \hat{e}^\beta_B\, \bar{D}(z)\, \mathsf{F}^{\rm GR}_{\alpha\beta},
\label{eq:bgo_GR_correction}
\end{equation}
with
\begin{equation}
\begin{split} 
\mathsf{F}^{\rm GR}_{\alpha\beta} \equiv \, 
&\tilde{h}^{(0)\gamma\zeta} \, \delta \mathcal{R}_{\zeta\alpha\beta\gamma}
+ \frac{1}{2} \partial_\gamma \delta \partial^\gamma \tensor{\tilde{h}}{^{(0)}_{\alpha\beta}}
- \frac{1}{2} \delta \tensor{\Gamma}{^\gamma_{\beta\zeta}} \, \partial^\zeta \tensor{\tilde{h}}{^{(0)}_{\alpha\gamma}}
- \frac{1}{2} \delta \tensor{\Gamma}{^\gamma_{\alpha\zeta}} \, \partial^\zeta \tensor{\tilde{h}}{^{(0)}_{\beta\gamma}} 
- \frac{1}{2} \tensor{\tilde{h}}{^{(0)}_{\beta\gamma}} \, \partial_\zeta \delta \tensor{\Gamma}{^\zeta_{\alpha\gamma}}+\\&
+ \frac{1}{2} \tensor{\tilde{h}}{^{(0)}_{\alpha\gamma}} \, \partial_\zeta \delta \tensor{\Gamma}{^\zeta_{\beta\gamma}}
+ \frac{1}{2} \delta \tensor{\Gamma}{^\gamma_{\gamma\zeta}} \, \partial^\zeta \tensor{\tilde{h}}{^{(0)}_{\alpha\beta}}
- \frac{1}{2} \delta \tensor{\Gamma}{^\gamma_{\beta\zeta}} \, \partial^\zeta \tensor{\tilde{h}}{^{(0)}_{\alpha\gamma}}
- \frac{1}{2} \delta \tensor{\Gamma}{^\gamma_{\alpha\zeta}} \, \partial^\zeta \tensor{\tilde{h}}{^{(0)}_{\beta\gamma}}.
\label{eq:F_GR}
\end{split}
\end{equation}
\end{widetext}
The integral in Eq.~\eqref{eq:bgo_GR_correction} has been split into the three distinct regions discussed above and illustrated in the right panel of Figure \ref{fig:propagation_panels}. The source term $\mathsf{F}^{\rm GR}_{\alpha\beta}$ encapsulates the full set of effective interactions contributing to bGO corrections in GR, evaluated at linear order in $\Psi$. We adopt the same notation for the function $\mathsf{F}_{\mu\nu}$ as in \cite{Dalang:2021qhu}, where bGO corrections are derived in GR for a point lens.

The analysis focuses on the $mm$ and $ll$ polarization modes, which are directly related (via a circular polarization basis \cite{Isi:2022mbx}) to the physical GWs polarizations $h_+$ and $h_\times$. 
We first consider the case in which GW propagate with an impact parameter small enough to intersect the lens, i.e., $b < R$. Explicitly evaluating the integrals in Eq.~\eqref{eq:bgo_GR_correction} in each region, using the appropriate expressions for the gravitational potential $\Psi$ provided in Eqs.~\eqref{eq:grav_potential_out} and~\eqref{eq:grav_potential_in}, one finds that the contributions from the exterior regions vanish identically, while the contribution in the interior of the lens yields a non-zero correction. Specifically, the integral over the interval $[-z_{\rm lens}, +z_{\rm lens}]$ can be computed analytically, yielding
\begin{equation}\label{GR dispersion equation}
\delta \tilde{\alpha}^{(1)}_{\circ}(z_o) \propto -\frac{4 GM}{fR^2}\sqrt{1-\frac{b^2}{R^2}}\, \tilde{\alpha}^{(0)}_{\circ}(z_s)\,
\end{equation}
where $\circ \in \{mm, ll\}$. This result shows that even within GR, a spherically symmetric extended lens can induce non-trivial dispersive corrections to the $mm$ and $ll$ modes, thereby generating a phase shift in the physical GW polarizations $h_+$ and $h_\times$.
By plugging Eq.~\eqref{GR dispersion equation} into Eq.~\eqref{perturbed amplitudes WFL new}, the aforementioned result can be rewritten in terms of the dispersion parameter defined in Eq.~\eqref{beta}, leading to
\begin{equation}\label{beta_only_GR}
\boxed{\beta_{\circ}=\frac{4(GM)^2}{R^2}\sqrt{1-\frac{b^2}{R^2}}}.
\end{equation}

Fig.~\ref{otherobjects} shows the general-relativistic LID for different astrophysical lenses.
We show results for extended objects, approximated as spherical and homogeneous distributions: a globular cluster ($M = 10^5\,M_\odot$, $R \sim 30~\rm pc$) and dwarf galaxy (with $M = 10^8\,M_\odot$, $R \sim 2~\rm kpc$).
Due to their large masses, these systems satisfy the short-wave condition $GMf < 1$ for ground detectors (both) and space-borne interferometers (dwarf galaxy), but their associated dispersive corrections are strongly suppressed.
Fig.~\ref{otherobjects} also shows results for near-field lenses, such as the Earth or the Moon, where LID corrections are much stronger. These objects violate the short-wave condition, but they cover a large fraction of the sky for detectors located on their surface. As we will argue below, this will allow $\sim 100\mathrm{Hz}$ interferometers on the Earth and the proposed $\sim 1\mathrm{Hz}$ Lunar Gravitational Wave Antenna (LGWA)~\cite{Ajith:2024mie} to probe LID in GR and beyond.

\begin{figure*}[t]
    \centering
    \includegraphics[width=\textwidth]{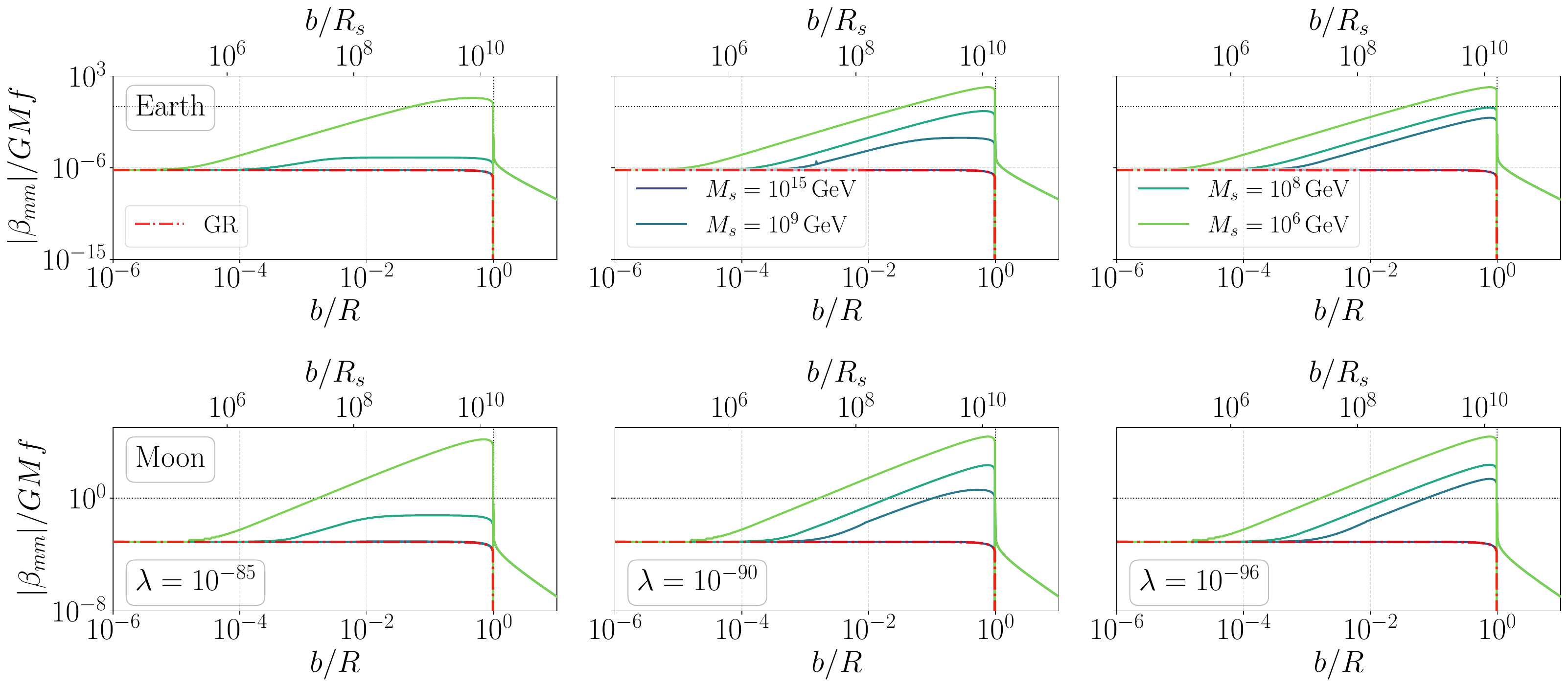}
    \caption{This plot shows the dispersion parameter $|\beta_{mm}/GMf|$ as a function of the impact parameter, normalized to the object's radius $b/R$, for two different lenses: Earth (top row, with reference frequency $f_{\rm GW}\sim 100$ Hz) and Moon (bottom row, with $f_{\rm GW}\sim 1$ Hz). Each column corresponds to a different value of the scalar field self-coupling $\lambda$. Different colors in the plot represent different values of the symmetron coupling to matter $M_s$, with the GR result shown as a red dash-dot line for reference. An additional black dotted line at $\beta_{mm}/GMf\sim1$ and a vertical line at $b/R\sim1$ are shown in all panels: the first indicates when dispersive effects become $\mathcal{O}(1)$ and the second indicates the boundary of the lens. On the top of each subplot, we show the impact parameter rescaled by the Schwarzschild radius of the corresponding lens, $b/R_s$.}
    \label{linear plot earth and moon}
\end{figure*}

Using the Earth as reference, Eq.~\eqref{beta_only_GR} can be recast as
\begin{equation}\label{beta GR}
   \frac{\beta_{\circ}}{GMf} \sim 3\times10^{-7} \left( \frac{M}{M_\oplus} \right) \left( \frac{R}{R_\oplus} \right)^{-2} \left( \frac{f}{100~\mathrm{Hz}} \right)^{-1}
\end{equation}
where the frequency dependence of LID has been reintroduced following Eq.~\eqref{beta}, assuming typical ground GW detector band.
The red dotted line in Fig.~\ref{linear plot earth and moon} shows that, in the case of the Earth, the resulting waveform phase shift is of order $\sim 10^{-6}$: a small but non-negligible correction that already signals the emergence of GW dispersion within GR. In contrast, the Moon leads to a larger effect: the lower frequency and the smaller lunar mass amplifies the frequency-dependent corrections through the $1/f$ factor in $\beta_{\circ}/(GMf)$, resulting in an enhancement of approximately two orders of magnitude compared to the Earth configuration.

In contrast, when the GW propagates entirely outside the lens ($b > R$), the signal never intersects the matter distribution, and the gravitational potential along its path remains identical to that of a point-like mass, i.e., Eq.~\eqref{eq:grav_potential_out}. In this configuration, the internal region of integration (Region~II) is effectively absent. As a result, the entire propagation occurs in vacuum, and there is no need to split the integral of Eq.~\eqref{BGO GR integration} into regions with distinct gravitational potentials. In such a scenario, the total integral vanishes and there are no dispersive corrections. This behavior is fully consistent with the results of  \cite{Dalang:2021qhu}, which showed that no bGO corrections arise in GR for the $mm$ and $ll$ modes. We conclude that the finite size and internal structure of the lens are responsible for sourcing dispersive effects.

\subsection{Symmetron gravity}\label{dispersive corrections in symmetron}
In this section we present the computation of bGO corrections in symmetron gravity and the resulting LID on the GW signal, adopting the Earth and the Moon as lenses, consistent with the analysis performed in GR. Unlike GR, the symmetron framework introduces additional parameters, such as $M_s$ and the self coupling $\lambda$, which control the profile and dynamics of the scalar field. The evaluation of dispersive corrections is more challenging than in GR, as the underlying propagation equation is more complex. In what follows, our aim is understanding how the interplay between the symmetron parameters and those associated with the lens (implicitly accounting for the effects of screening) can modulate and shape the resulting bGO corrections.

The formal solution for the bGO corrections in the symmetron case is given by Eq.~\eqref{bGO formal symmetron solution}. Assuming the GW geodesics intersecting the lens as illustrated in Figure \ref{fig:propagation_panels}, the corresponding bGO integral can be evaluated along the $\hat{z}$-axis and split into three regions, following the same procedure adopted in GR in Eq.~\eqref{eq:bgo_GR_correction}. For the symmetron, the expression reads
\begin{widetext}
    \begin{equation}\label{bGO symmetron in spherical lens}
\tilde{\alpha}_{AB}^{(1)}(z_o) = 
 i \frac{A[\phi(z_o)]}{D(z_o)} \left( \int_{-\infty}^{-z_{\rm lens}} + \int_{-z_{\rm lens}}^{+z_{\rm lens}} + \int_{+z_{\rm lens}}^{+\infty} \right) \frac{dz}{\Omega} \, \hat{e}^\mu_A \hat{e}^\nu_B A[\phi(z)] D(z)\mathsf{F}^{\rm SY}_{\mu\nu},
\end{equation}
where
\begin{equation}\label{bgo symmetron linearized}
\begin{split}
\mathsf{F}^{\rm SY}_{\mu\nu} &\equiv \,
- \frac{ \delta \tilde{h}^{(0)}_{\alpha\beta} \, V(\phi) }{ M_{\rm P}^2\, A(\phi)^3 }
- \frac{ \tilde{h}^{(0)}_{\alpha\gamma} \, \delta \mathcal{R}_\beta^\gamma }{ A(\phi) }
+ \frac{ \tilde{h}^{(0)}_{\alpha\beta} \, \delta \mathcal{R} }{ 2\, A(\phi) }
- \frac{ \tilde{h}^{(0)\,\gamma\zeta} \, \delta \mathcal{R}_{\zeta\alpha\beta\gamma}}{ A(\phi) }
+ \frac{ \Box \delta \tilde{h}^{(0)}_{\alpha\beta} }{ 2\, A(\phi) }
- \frac{ k(\phi)\, \delta \tilde{h}^{(0)}_{\alpha\beta} \, A(\phi) \, \phi_\gamma \, \phi^\gamma }{ 2\, M_{\rm P}^2 }
- \frac{ \delta \tensor{\Gamma}{^\gamma_\beta_\zeta} \, \partial^\zeta \tilde{h}^{(0)}_{\alpha\gamma} }{ 2\, A(\phi) } +\\
&\quad
- \frac{ \delta \tensor{\Gamma}{^\gamma_\alpha_\zeta} \, \partial^\zeta \tilde{h}^{(0)}_{\beta\gamma} }{ 2\, A(\phi) }
- \frac{ \tilde{h}^{(0)}_{\beta\gamma} \, \partial_\zeta \delta \tensor{\Gamma}{^\zeta_\alpha_\gamma} }{ 2\, A(\phi) }
- \frac{ \tilde{h}^{(0)}_{\alpha\gamma} \, \partial_\zeta \delta \tensor{\Gamma}{^\zeta_\beta_\gamma} }{ 2\, A(\phi) } 
+ \frac{ \delta \tensor{\Gamma}{^\gamma_\gamma_\zeta} \, \partial^\zeta \tilde{h}^{(0)}_{\alpha\beta} }{ 2\, A(\phi) }
+ \frac{ \tilde{h}^{(0)}_{\alpha\gamma} \, \partial_\zeta \delta \tensor{\Gamma}{^\zeta_\beta_\gamma} }{ 2\, A(\phi) }
+ \frac{ 2\, \delta \tilde{h}^{(0)}_{\alpha\beta} \, \phi_\gamma\, \phi^\gamma\, A'(\phi) }{ A(\phi)^2 }+ \\
&\quad
+ \frac{ \tilde{h}^{(0)}_{\beta\zeta} \, \delta \tensor{\Gamma}{^\xi_\alpha_\beta} \, \phi^\gamma\, A'(\phi) }{ A(\phi)^2 }
+ \frac{ \tilde{h}^{(0)}_{\alpha\zeta} \, \delta \tensor{\Gamma}{^\xi_\beta_\zeta} \, \phi^\gamma\, A'(\phi) }{ A(\phi)^2 }
+ \frac{ 2\, \tilde{h}^{(0)}_{\gamma\zeta} \, \delta \tensor{\Gamma}{^\xi_\alpha_\beta} \, \phi^\gamma\, A'(\phi) }{ A(\phi)^2 }
+ \frac{ \delta \tensor{\Gamma}{^\xi_\gamma_\zeta} \, \phi^\gamma\, A'(\phi)\, \tilde{h}^{(0)}_{\alpha\beta} }{ A(\phi)^2 } +\\
&\quad
+ \frac{ 2\, \partial_\beta \delta \tilde{h}^{(0)}_{\alpha\gamma} \, \phi^\gamma\, A'(\phi) }{ A(\phi)^2 }
- \frac{ \partial_\alpha \delta \tilde{h}^{(0)}_{\alpha\beta} \, \phi^\gamma\, \phi_\gamma\, A'(\phi)^2 }{ A(\phi)^3 }
+ \frac{ 2\, \delta \tilde{h}^{(0)}_{\alpha\beta} \, \phi_\gamma\, \phi^\gamma\, A''(\phi) }{ A(\phi)^2 }.
\end{split}
\end{equation}
\end{widetext}
The expression~\eqref{bgo symmetron linearized} is derived by taking the operator in Eq.~\eqref{eq:propagator_operator} and linearizing to first order in $\Psi$ the functions $\tensor{\mathsf{K}}{_\mu_\nu^\alpha^\beta^\gamma^\rho}$, $\tensor{\mathsf{A}}{_{\mu\nu}^{\alpha\beta\gamma}}$, and $\tensor{\mathsf{M}}{_{\mu\nu}^{\alpha\beta}}$, whose explicit forms are given in Appendix~\ref{appendix:symmetron_functions}. Here, the superscript SY denotes that the quantity is referred to symmetron gravity. As a consistency check, we confirm that the above expression reproduces the GR limit shown in Eq.~\eqref{eq:F_GR}, taking $A(\phi) \to 1$ (or, equivalently, $M_s \to \infty$) and $\phi \to 0$. This expression also reduces to the Brans-Dicke case investigated in \cite{Menadeo:2024uoq} if we choose $A(\phi)=1/\sqrt{\phi}$, which reproduces the standard Brans-Dicke Lagrangian yielding
Eq.~101 in \cite{Menadeo:2024uoq}.

At this stage, Eq.~\eqref{bGO symmetron in spherical lens} can be explicitly evaluated by computing all relevant quantities using the background metric~\eqref{wfl metric}, together with the scalar field profiles~\eqref{phi in} and~\eqref{phi out} presented in Sec.~\ref{sec:Scalar field solutions}, and the corresponding gravitational potentials~\eqref{eq:grav_potential_in} and~\eqref{eq:grav_potential_out}. These expressions are employed separately in the inside and outside regions of the lens, depending on the integration domain. Unlike the GR case discussed in Sec.~\ref{dispersive corrections in GR}, the integration in the symmetron framework is performed numerically.
\subsubsection{Numerical results}\label{numerical results}

This section presents and discusses the results obtained from the integration of Eq.~\eqref{BGO GR integration}. As stated in Sec.~\ref{dispersive corrections in GR}, the analysis focuses exclusively on the modes $\tilde{\alpha}_{mm}^{(1)}$ and $\tilde{\alpha}_{ll}^{(1)}$, whose associated phase shifts are encoded in the parameter $\beta_{\circ}/(GMf)$, as discussed in the general framework in Sec.~\ref{General formalism for GW propagation}. All the unconstrained mode components from Eq.~\eqref{unconstrained modes}, such as $mn$ and $nl$ vanish identically. An additional scalar-induced mode, corresponding to the $nn$ component, also appears, but it is irrelevant for our purposes and has been discussed in \cite{Menadeo:2024uoq, Cusin:2019rmt}.

The dispersive corrections are shown in Figure \ref{linear plot earth and moon}, displaying the behavior of $\beta_{mm}/(GMf)$ (behaviour identically mirrored for the $mm$-mode) as a function of the impact parameter $b$, normalized by the lens radius $R$ (the behavior of \textit{ll} is identically the same). The three panels correspond to different values of the symmetron parameter $\lambda$ (down to its lower bound, according to Eq.~\eqref{lambda max}), while each curve within a panel refers to a different choice of the coupling scale $M_s$. For $M_s = 10^{15}~\mathrm{GeV}$, the correction becomes indistinguishable from that of GR; this value of $M_s$ effectively represents the GR limit $M_s \to \infty$.
\begin{figure*}[t]
    \centering
    \includegraphics[width=\textwidth]{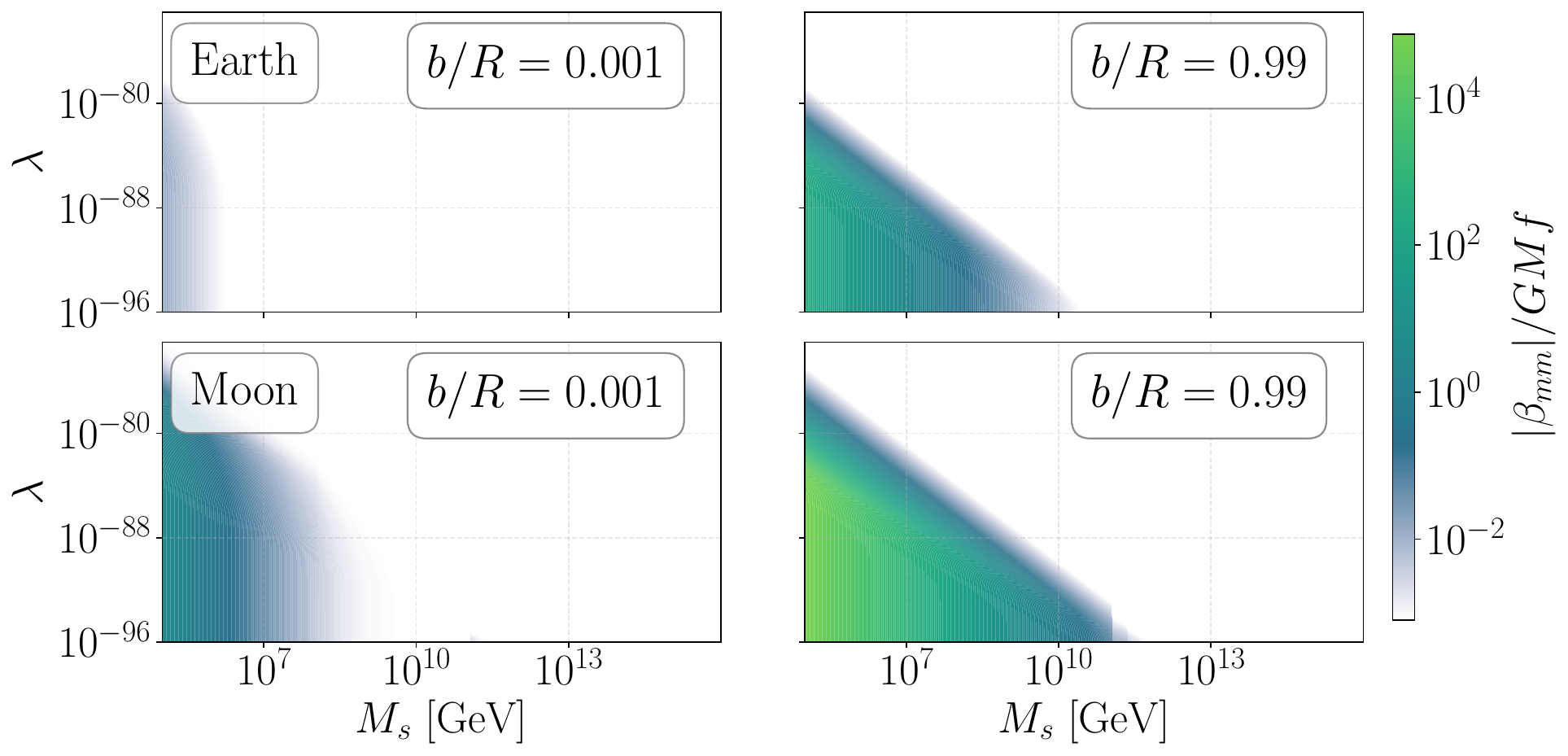}
    \caption{Contour plot of $\beta_{mm/ll}/(GMf)$ in the allowed symmetron parameter space $(M_s,\lambda)$. Each panel corresponds to a different lens: Earth (top row, $f_{\rm GW}\sim 100$ Hz) and Moon (bottom row, $f_{\rm GW}\sim 1$ Hz) for fixed impact parameters $b/R\sim10^{-3}$ (left column) and $\sim0.99$ (right column), respectively. The fading of the contours toward the white region indicates the GR limit.}
    \label{fig:Contour_plot}
\end{figure*}

Although the Earth and Moon do not satisfy the short-wave condition $GM f \gtrsim 1$, we can use our framework to anticipate dramatic departures from GW propagation in vaccuum. As we will argue in Sec.~\ref{sec:critical_freq_constraints}, a large LID parameter $\beta \gtrsim 1$ signals that the effective mass for GWs is large enough that propagation through the Earth/Moon becomes exponentially suppressed. Because the detectors are placed directly on the surface (or slightly underneath), the object blocks a large portion of the sky, creating a large blind spot or at least modifying the antenna pattern significantly.

Figure~\ref{linear plot earth and moon} shows the existence of symmetron parameters where GW dispersion is strong. 
The distribution on the symmetron parameter space shows a similar behaviour for both lenses, but enhanced in the lunar case; this amplification (as we already discussed in GR) is primarily driven by the lower frequency adopted for the lunar configuration, which naturally increases the value of $\beta_{\circ}/GMf$ and thus makes dispersion stronger. 
In addition, the intrinsic properties of the lens (its mass and radius) also modulate the strength of the dispersion. Since the GW phase term scales inversely with the lens mass, the smaller mass of the Moon (compared to the Earth) further contributes to enhancing the overall dispersive effect. Moreover, the Moon shallower gravitational potential varies more gradually across the propagation path, producing a more diffuse transition between the screened and unscreened regions near the lens boundary. This more extended variation of the metric gets accumulated in the phase correction, effectively amplifying the dispersive response. Conversely, the Earth’s stronger gravitational potential yields sharper curvature confined to a smaller region.

Increasing $M_s$ leads to a stronger suppression of the scalar field, making the modified gravity contributions negligible and effectively recovering the GR prediction. In contrast, decreasing $M_s$ enhances the scalar coupling (via the conformal factor $A(\phi)$), resulting in significantly larger dispersive effects. These corrections are particularly prominent near the boundary of the lens, where the scalar field develops steep gradients. 

Decreasing the self-coupling $\lambda$
amplifies the overall dispersive corrections across all configurations. For small values of $M_s$, the amplitude of the effect is already strong at $\lambda = 10^{-85}$ and exhibits only mild changes for lower $\lambda$, indicating an early saturation due to the strong conformal coupling given by $A(\phi)$. In contrast, larger $M_s$ values show a more pronounced enhancement as $\lambda$ decreases.

Analogously to the GR case, for $b < R$, the entire contribution to the bGO correction originates from Region~II. As the impact parameter approaches the lens radius ($b \to R$), the inner integration domain progressively shrinks, reducing the influence of this region. However, in this near-boundary regime, the scalar field undergoes a transition across the thin shell, where its profile and its derivatives bridge between the screened and unscreened configurations. As a result, dispersive effects tend to peak in this region, just before vanishing. This behavior explains the sharp features observed close to the edge of the lens, where the correction rises and then rapidly drops to zero precisely at $b = R$, when Region~II disappears and the GW trajectory no longer intersects the lens.

In the limit $b \ll R$ (or $b\to0$), the geodesic penetrates deeper into the core of the lens, where the scalar field profile is nearly constant and close to zero (as shown in the left panel of Figure \ref{fig:scalar_field}). Since Eq.~\eqref{bgo symmetron linearized} depends not only on the field itself but also on its first and second derivatives, the overall suppression of $\phi$ and its gradients leads to a strong reduction of the dispersive corrections. As a result, $\beta_{\circ}/(GMf)$ gradually converges to a plateau that asymptotically approaches the GR result. This behavior highlights the fact that, once the scalar field is sufficiently suppressed (as is the case deep inside the screened core) any modification to GR becomes observationally indistinguishable. Regardless of the underlying model parameters or the specific lens properties, the key factor is the location of the geodesic itself: only trajectories intersecting regions where the scalar field profile gives a strong contribution are able to probe departures from standard gravitational dynamics.

Let us now comment on the discontinuity visible at \( b = R \), where the signal transitions from a trajectory intersecting the lens to one that remains entirely outside. This feature arises directly from the physical setup of the problem, in which the signal propagates through a piecewise medium composed of vacuum (Region I), constant density matter (Region II), and vacuum again (Region III). Since the lens density profile is modeled as a step function, the corresponding gravitational potential \(\Psi(r)\), obtained by solving the Poisson equation, is continuous across the boundary along with its first derivative, but displays a discontinuity in its second derivative. The bGO corrections in Eq.~\eqref{bgo symmetron linearized} involve geometric quantities such as linearized Christoffel symbols and curvature tensors, which, by definition, are proportional to first and second derivatives of the gravitational potential \(\Psi\), respectively 
The bGO corrections thus depend explicitly on second derivatives of \(\Psi\), and therefore inherit any discontinuity present in them. 

This discontinuity could be avoided by adopting a continuous density profile for the lens across the full propagation domain, so that no sharp transition between vacuum and matter occurs. However, since this would entail using a fully numerical profile instead of the analytical ones provided in \eqref{phi in} and \eqref{phi out}, we neglect this possibility.

For \( b > R \), the signal propagates entirely outside the lens. In this case, only small residual contributions remain, originating from Regions~I and~III. These are observable as decaying tails, but are significantly suppressed compared to the dominant corrections accumulated when the geodesic intersects the dense core of the lens.

An additional visualization of LID effects in symmetron gravity is provided by Figure \ref{fig:Contour_plot}, where the parameter $\beta_{\circ}/(GMf)$ is evaluated across the entire $(M_s, \lambda)$ parameter space, for fixed values of the impact parameter $b/R$ and frequency. We focus on two representative regimes: $b/R = 10^{-3}$, corresponding to geodesics deeply embedded within the lens, and $b/R = 0.99$, near the boundary, where dispersive phenomena tend to peak.
The contour plots also provide a clear visualization of the screening mechanism at play. For geodesics deep inside the lens ($b/R = 10^{-3}$), significant phase shifts only appear in regions characterized by strong coupling, i.e., small $M_s$ and $\lambda$. As $M_s$ increases, the scalar field becomes progressively suppressed, and the corrections rapidly fade in the GR limit indicated by the white region. Conversely, for near-boundary propagation ($b/R = 0.99$), the scalar field gradients are stronger, and sizable corrections persist over a much broader range of parameter space, with $\beta_{mm/ll}/(GMf)$ reaching, and in some cases exceeding, unity. These transition regions, where the scalar field is activated, are thus the primary sources of large LID effects. The investigation is carried out by scanning over a range of values for the $M_s$, varying approximately from $10^5~\mathrm{GeV}$ to $10^{15}~\mathrm{GeV}$, consistent with current observational and experimental constraints on symmetron by laboratory tests and astrophysics, as discussed in \cite{Burrage:2017qrf} and shown in Figure \ref{fig:constraint_critical_freq}.

We present results on the $M_s-\lambda$ plane, assuming that the scalar field mass is linked to the onset of dark energy (cf.~Eq.~\eqref{eq:mu from Ms} and Ref.~\cite{Hinterbichler:2011ca}), reducing the symmetron free parameters from three to two.
However, we made sure that this is compatible with the other assumptions in our analysis. Specifically, the Earth (in addition to the Sun and the Milky Way) is screened for almost the entire parameter range we consider (meaning that $\alpha$ in \eqref{alpha} is $>1$ for $10^{5}<M_s<10^{15} \rm\,GeV$), with the exception of the tail end of larger $M_s$. In that case, however, dispersive effects are already irrelevant, as shown in Figures \ref{fig:Contour_plot} and Figure \ref{fig:constraint_critical_freq}. It is interesting to note that the Earth being screened does not prevent it from exhibiting dispersive effects, in accordance with the presence of such effects in GR with an extended lens as well (see Section \ref{dispersive corrections in GR}). The fact that the Milky Way is screened in this parameter range does not make dispersive effects unobservable, since our analysis does not extend to galactic, or even Solar System scales: we are only concerned with effects only in the vicinity of the Earth itself.

Most conservatively, we require a separation of scales, such that the scalar sourced by the Sun and the Earth are decoupled.
This translates to a range of $\mu$ such that
$ R_{\rm Earth}<\mu^{-1}\lesssim 1 \text{AU}.
$ 
The upper limit, $1{\rm AU}\sim 10^{11}\rm\,m$ is around the same order of magnitude of $\mu^{-1}$ (the symmetron force range) for $M_s\sim 10^5 \rm\,GeV$. Therefore, our results are reliable in the lower range of $M_s$, precisely when the dispersion effects are most relevant, which validates our analysis. Values of $M_s$ 
closer to the upper limit of the considered range do not satisfy $\mu^{-1}\lesssim 1 {\rm AU}$,  
but dispersive effects become irrelevant there anyway. Astrophysical tests start at those values of $M_s$ because the Milky Way starts to become unscreened.
Note that the condition on $\mu$ can be relaxed if the symmetron is not related to the onset of dark energy, i.e.~departing from Eq.~\eqref{eq:mu from Ms}. Extending the range for $\mu$ does not necessarily exclude large LID effects, but the background value of the scalar field (around the Earth, as sourced by some other object) needs to be considered, e.g. as a boundary condition.
Situations in which the lens can not be considered as isolated will likely introduce additional signatures that can be used do characterize LID and the underlying theory.
\footnote{For instance, an angular dependence will be introduced if the scalar field on the Earth is influenced by a third object (e.g. the Sun, Moon, Milky Way). Then, LID effects will not only depend on the relative orientation of the source, Earth and detector, but also the direction of the third body at the time of the detection.}

\section{Critical frequency and GW blocking by near-field lenses}\label{sec:critical_freq_constraints}

\begin{figure}
    \centering
    \includegraphics[width=\linewidth]{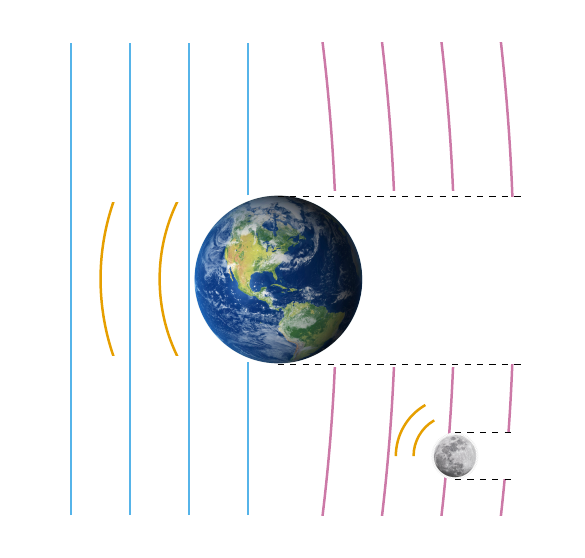}
    \caption{Schematic diagram of GW blocking and reflection by extreme dispersion, in the case of near-field lenses. Incoming plane GWs (blue) partially propagate (purple) around the Earth, but are also reflected (orange) by the Earth and the Moon.}
    \label{blocking}
\end{figure}

{\justifying 
In this section, we take the consequences of lens-induced dispersion to the extreme, considering the Earth and the Moon as lenses. For almost the whole range of $M_s$ considered in this work, the Earth is a screened object (the screening parameter $\alpha\gg 1$), except for the extreme lower end of the parameter range ($M_s\sim 10^{5}\, \rm GeV$). This case represents the situation where dispersion is strongest and might block GW propagation completely. This allows us to find the observational constraints on the symmetron parameters derived from GW dispersion.

Very strong dispersion can prevent GW propagation when the wave frequency becomes similar to (or smaller than) the effective mass of the $+,\times$ fields inside the astrophysical body (lens),
given by $\tensor{{\mathsf{M}}}{_{\mu\nu}^{\alpha\beta}}$ in the propagation equation \eqref{eq:propagator_operator}. This effect is similar to that of introducing a massive graviton, which modifies the waves' dispersion relation, but now the field's mass depends on the space-time position and propagation direction. To avoid this complication and the need to expand the formalism beyond the usual approximations (short-wave, paraxial), we will present a heuristic discussion of the effect, leaving a detailed analysis for future works.

GW blocking is similar to the behavior of electromagnetic waves in the Earth’s ionosphere \cite{PhysRev.126.1899}. There, for entirely different physical reasons than those in our problem, the ionosphere acts as a cutoff medium: radio waves with wavelengths longer than about (typically in the range $~1-10~\mathrm{MHz}$, corresponding to wavelengths above $~30\mathrm{m}$) generally cannot propagate through the plasma. Instead, they are reflected back or strongly attenuated, because the free electrons in the ionosphere oscillate at a natural frequency. Waves below this threshold, known as the atmosphere's \textit{plasma frequency}, are unable to propagate and are either absorbed or reflected. By analogy, we will refer to the Earth's \textit{critical LID frequency} for GWs.

We are interested in the case in which the Earth as a gravitational lens causes dispersion that obstructs GW propagation. To illustrate this, we can introduce a simplified toy model: a scalar wave described by a Klein-Gordon field (although the analysis in the previous sections is generic and encompasses a much broader phenomenology than the specific case treated here). The scalar field equation reads
\begin{equation}
\Box\phi+m^2\phi^2=0,
\end{equation}
with 
\begin{equation}
\phi(x,t)=e^{-i f t}g(x),
\end{equation}
encountering a one-dimensional potential barrier, where the potential takes the form of a simple mass term. The portion of the wave passing through the barrier experiences a phase shift, which can be equated to the lens-induced phase shift $\beta$ defined in \eqref{beta}, characterizing the strength of GW dispersion on the waveform.
In this simple set-up, the freely propagating wave (before encountering the barrier, as indicated by the subscript $_1$) has wave vector $k_1=f$, while the wave going through the barrier has wave vector $k_2=\sqrt{f^2-m^2}$.
General solutions have the following structure
\begin{equation}
\begin{aligned}
g_{1}(x)&=e^{i k_1 x}+R e^{-i k_1+x} \\
g_2(x)&=T e^{-i k_2 x}, 
\end{aligned}
\end{equation}
where $R$ and $T$ are the reflection and transmission coefficients, respectively, found through imposing the boundary conditions ensuring continuity of the function and its derivatives at the boundary of the barrier. The first term in $f_1(x)$ is the incident wave and the decaying solution for $f_2(x)$ is valid when $f<m$. The boundary conditions impose
\begin{equation}
    1+R=T \quad\quad k_1(1-R)=k_2T,
\end{equation}
yielding the coefficients
\begin{equation}
\label{coeffs}
    R=\dfrac{k_1-k_2}{k_1+k_2} \quad\quad T=\dfrac{2 k_1}{k_1+k_2}.
\end{equation}

Given $k_2=\sqrt{f^2-m^2}$, there are two limits: 
\begin{itemize}
    \item In the low-frequency limit, $f\ll m$, and we can expand in the small parameter $\left|{f/m}\right|$ to obtain
    \begin{equation}
        k_2=i m\left(1-\left|\dfrac{f}{m}\right|^2+...\right)
    \end{equation}
    \item In the high-frequency limit, $f\gg m$, and we can expand in the small parameter $\left|{m/f}\right|$ to obtain
    \begin{equation}
    \label{k2}
        k_2=f\left(1-\left|\dfrac{m}{f}\right|^2+...\right).
    \end{equation}
\end{itemize}
In the low-frequency limit, since $k_2$ is imaginary, we can denote $k_2 = i\kappa$, so that $\kappa =\sqrt{m^2-f^2}>0$. Using \eqref{coeffs}, the transmission coefficient is therefore
\begin{equation}
    T_{f\ll m}=\frac{2f}{f+i\kappa},
\end{equation}
and the amplitude of the transmitted wave is
\begin{equation}
    |T_{f\ll m}|=\frac{2f}{\sqrt{f^2+\kappa^2}}\approx\frac{2f}{m}.
\end{equation}
Its phase is the argument of the complex quantity $T_{f\ll m}$, namely
\begin{equation}
    \arg(T_{f\ll m})=-\arctan\left(\frac{\kappa}{f}\right).
\end{equation}

The amplitude of signals with $f<f_c$ decays exponentially inside the massive region: the signal is then blocked by the object, which acts as a reflector.

In the high-frequency case, the amplitude of the transmitted wave, from \eqref{coeffs}, is simply
\begin{equation}
    |T_{f\gg m}|=\frac{2f}{f+\sqrt{f^2-m^2}}\approx 1,
\end{equation}
and the wave is fully transmitted.

In the regime $f\gg m$, the results can be derived by a short wave expansion, which converges to the exact result of the 1-D barrier.

\begin{figure*}[t]
    \centering
    \includegraphics[width=0.7\textwidth]{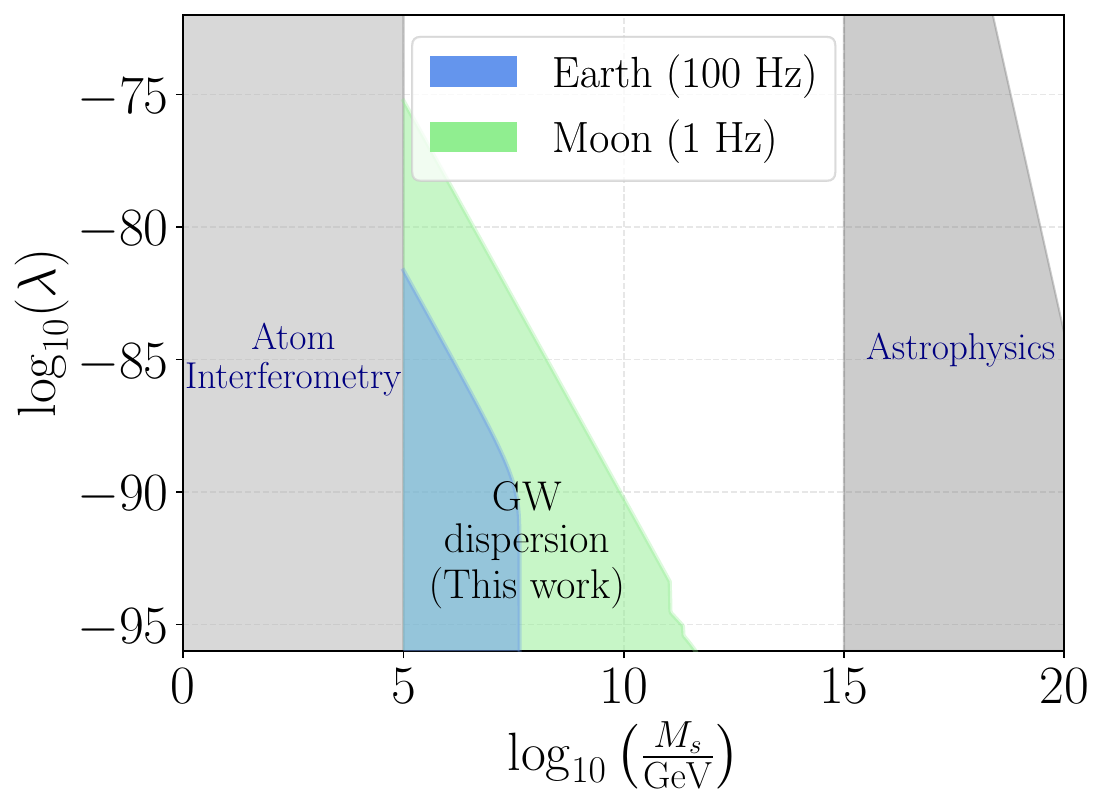}
    \caption{ 
    Regions of the symmetron parameter space where GWs are blocked by the bodies where the detector lays, e.g. $f<f_c$ for typical frequencies of existing/planned ground detectors and a lunar interferometer at lower frequencies. The shaded regions can be excluded by the observation of sources propagating through the body or strong modifications on the detector response function. 
    }
    \label{fig:constraint_critical_freq}
\end{figure*}

We will now match the results of the 1D potential barrier to the LID calculation to estimate the critical frequency for a given $\beta$. To this end, we need to compare both results in the high-frequency limit ($f\gg m$), in which both methods converge.
From \eqref{k2}, we can find the change in wave vector
\begin{equation}
\Delta k_2=f-k_2\approx\dfrac{m^2}{f} \quad (f\gg m)\,.
\end{equation}
The phase of the transmitted wave can be written as
\begin{equation}
\Delta \varphi=\Delta k_2 \Delta z,
\end{equation}
where $\Delta z$ is the length of the barrier.
For the 1-D barrier, the critical frequency is given by the field's mass $f_c = m$,
since waves with $f\sim m$ will be exponentially blocked.
Substituting $\Delta k_2$, $m\sim f_c$ and equating $\Delta \varphi$ to the phase shift in the waveform $\dfrac{\tilde{\alpha}^{(1)}_{AB}}{\tilde{\alpha}^{(0)}_{AB}}$ in \eqref{betabox}, we find
\begin{equation}
\label{critfreq}
    \boxed{f_c\approx\sqrt{\dfrac{\beta_{\circ}}{GM\Delta z}}},
\end{equation}
where now $\Delta z \simeq 2\sqrt{R^2-b^2}$. For high frequencies $f\gg f_c$, the LID computation holds, while $f\lesssim f_c$ will produce substantial deviations, with potentially observable consequences.

Figure \ref{fig:constraint_critical_freq} shows the symmetron parameter space, where the Earth (Moon) fulfills the condition $f_c> 100\, (1)$ Hz. In this case, GWs with $f<f_c$ would be blocked by the Earth because of extreme LID, leading to novel potential constraints of symmetron gravity.
It is remarkable that the constraints in Figure \ref{fig:constraint_critical_freq} fall precisely within the ``desert'' between the laboratory (\cite{Elder:2016yxm, Burrage:2017qrf, Elder:2019yyp}) and astrophysical constraints, that neither is able to access. To the best of our knowledge, this region of parameter space can only be probed by space-based tests of Shapiro time delay \cite{Sakstein:2017pqi}, 
still at the proposal or planning stage. Not only the potential constraints derived from LID are novel and complementary to existing ones, but they are also completely independent from them, since they are derived from GW propagation (see Figure 10 of \cite{Burrage:2017qrf} for a compendium of existing constraints before the present work). While our results have been derived by considering only the strongest LID configurations, this phenomenon is very general, as it arises in any non-minimal theory of gravity with an effective mass term.

Although simple, this 1-dimensional analogy illustrates the role of an effective mass, but several factors need to be considered in a future, detailed calculation. 
Rather than 1D, the Earth is approximately spherical: even if blocked in its interior, low frequency GWs can be diffracted around its surface. This may lead to a potentially observable signal even for sources located towards the antipodes of the detector, but it will drastically change the antenna pattern. 
For distant lenses, the astrophysical object becomes a reflector of GWs: this effect will cause an additional repetition of the signal, with a delay given by the additional travel time, sky localization consistent with the object and amplitude set by the ratio of the object's size to its distance
\footnote{For instance, a signal reflected by the Moon would produce a repeated signal with delay $\sim2.6~\rm s$ and amplitude ratio $\sim \sqrt{(2\pi R_{\rm Moon}^2)/(4\pi D_{\rm Earth/Moon}^2)}\sim0.002$. While faint, these echoes will exist for $\sim50\%$ of GW sources, and evidence can be found by combining many GW events~\cite{Zumalacarregui:2024ocb}.}.
A detailed calculation needs to account for the spherical lens, fully incorporate wave-optics at finite frequency and consider large deflection angles (e.g. no paraxial approximation). Such a wave-optics framework beyond GR will be challenging to derive, but its need is warranted by the potential to constrain non-minimal theories in general, in a manner complementary to astrophysical, cosmological and laboratory tests.

}
\section{Conclusions and outlook}\label{Conclusion}

In this work we have developed a general framework to study gravitational-wave (GW) propagation beyond geometric optics in inhomogeneous spacetimes, within and beyond General Relativity (GR). Building on the short-wave expansion formalism, we extended previous analyses of point-mass lenses to extended spherical matter distributions and generalized the approach to scalar–tensor theories with screening mechanisms. In particular, we computed for the first time the lens-induced dispersion (LID) of GWs caused by a homogeneous spherical lens in GR and in symmetron gravity.

The main results of this work are: 
\begin{itemize}
\item Dispersive effects are generic in the presence of matter inhomogeneities. Even in GR, propagation through the Earth causes LID at the level of $\sim 10^{-6}\left(\frac{f}{100{\rm Hz}}\right)$ Eq.~\eqref{beta GR}. Improved detector sensitivity may allow the detection this small GR correction by combining a large sample of binary mergers.

\item LID is significantly enhanced for screened modified gravity. In the case of the symmetron, the phasing parameter for a signal crossing a homogeneous object can be up to six to seven orders of magnitude larger than for Brans-Dicke, depending on the combination of theory parameters $(M_s, \lambda)$, see Figure \ref{linear plot earth and moon}. LID in symmetron is maximized close to the object's surface $b \sim R \gg GM$, while for Brans-Dicke it decays as $(GM/b)^3$. 

\item Dispersive phenomena may block GW propagation when the effective mass of $h_+,h_\times$ inside the lens becomes larger than the signal's frequency (below the \textit{critical LID frequency}, \eqref{critfreq}). In those cases, the lens becomes a reflector of GWs.

\item Considering the Earth/Moon as the lens leads to strong LID phasings in symmetron gravity, including regions in $M_s, \lambda$ where GW propagation is blocked, as shown in Figure \ref{fig:constraint_critical_freq}. This dramatically changes the way GWs are observed, and allows a direct test of the theory in regimes inaccessible to existing methods.
\end{itemize}

The universality of dispersive phenomena makes them a promising test of gravity. LID can be tested on any GW event, regardless of source properties, redshift information or an electromagnetic counterpart.
LID probes directly the non-derivative interactions of between the metric and new gravitational fields. These interactions are present in any non-minimal theory, and complement other effects such as lens-induced birefringence~\cite{Goyal:2023uvm} that require kinetic interactions, present in a subset of modified theories. 
Finally, at least in some cases LID can produce strong imprints and dramatic deviations, such as blocking and reflecting GWs.

The framework presented in this work can be applied to any Horndeski theory with luminal scalar waves, including theories related to the symmetron such as chameleon \cite{Khoury:2003rn} and $f(R)$ gravity \cite{Hu:2007nk}. No modification is needed for objects massive enough for the short-wave expansion to be valid: in this case, dispersive corrections to the phase can be integrated into a (frequency-dependent) lens equation. From this, one can compute the range of predictions expected in a realistic Universe, including the probabilistic distribution of phases, given a matter distribution, the source redshift, and the theory parameters. This program can be used to derive constraints for a theory, along the lines of \cite{Goyal:2023uvm}.

An important extension pertains theories with non-luminal scalar propagation: examples include coupled K-essence, Kinetic Gravity Braiding and Galileons. These theories involve rich kinetic, non-linear interactions and produce novel phenomena, such as the Vainshtein screening mechanism and self-accelerating solutions. Beyond scalar fields, the universality of LID in non-minimal theories suggests novel tests of vector-tensor theories, massive gravities \cite{Hinterbichler:2011tt, deRham:2014zqa} and relativistic alternatives to dark-matter \cite{Bruneton:2007si, Skordis:2020eui}.

Undoubtedly, the most promising manifestation of LID is the possibility that the Earth and Moon can become partial GW shields and reflectors. This phenomenon can be used to constraint theories in ways complementary to local, astrophysical and cosmological tests, but its development requires new theoretical tools. To understand this prediction, it is essential to incorporate diffractive effects beyond the usual approximations, in the regime in which the lens and observer are in close proximity. Even in the case that GWs are not strictly blocked, understanding this regime will dramatically expand the applicability of LID tests of gravity: from the few events aligned with a lens along the trajectory, to the $\mathcal{O}(0.5)$ fraction of GWs that cross the Earth. An exciting outcome of this program is the possibility of establishing the LID prediction in GR, even if it requires $\sim 10^5\text{--}10^6$ events observed by next-generation observatories at their exquisite sensitivities.

The most ambitious prospect of GW propagation over inhomogeneous space-times is the possibility to discover and characterize new gravitational degrees of freedom. The sensitivity of LID to cosmological modifications of gravity makes it a promising tool to study the dynamics of dark energy. 
Increasingly precise cosmological observations are beginning to reveal cracks in the cosmological constant paradigm~\cite{DESI:2024mwx, DESI:2025zgx, Rubin:2023jdq, DES:2024jxu} 
 
and favor non-minimal gravity theories~\cite{Ye:2024ywg,Wolf:2025jed,Berti:2025phi,Wolf:2025acj}. Theoretical advances in GW propagation beyond GR may provide an arsenal complementary to large-scale structure surveys and other cosmological probes.

\subsection*{Acknowledgments}
The authors thank Bruce Allen, Giulia Cusin, Charles Dalang, Arnab Dhani, Ben Elder, Jose María Ezquiaga, Pierre Fleury, Srashti Goyal, Macarena Lagos, Alberto Mangiagli, Shinji Mukohyama, Iggy Sawicki and Héctor Villarrubia-Rojo for interesting discussions related to this work. 

N.M. thanks Prof.~Capozziello for the support and acknowledges the Scuola Superiore Meridionale for sponsoring his period at the Albert Einstein Institute (AEI) in Potsdam. 
\appendix

\section{Symmetron background functions}\label{appendix:symmetron_functions}
In this appendix, we provide the functions  $\tensor{\mathsf{K}}{_\mu_\nu^\alpha^\beta^\gamma^\rho}$, $\tensor{\mathsf{A}}{_{\mu\nu}^{\alpha\beta\gamma}}$, and $\tensor{\mathsf{M}}{_{\mu\nu}^{\alpha\beta}}$, appearing in Eq.~\eqref{eq:propagator_operator} for symmetron gravity. They read
\label{appendix1}
\begin{widetext}
    \begin{align}
\label{BD effective metric tensor}
\tensor{\mathsf{{K}}}{_\mu_\nu^\alpha^\beta^\gamma^\rho}&\propto-\frac{1}{2 A(\phi)^2}\delta^{\alpha}_{\mu}\delta^{\beta}_{\nu}g^{\gamma\rho},\\ 
\label{eq: A BD}
\tensor{\mathsf{A}}{_\mu_\nu^\alpha^\beta^\gamma} 
&\propto - \frac{A'(\phi)}{A(\phi)^3}
\left( 2\, \phi^\alpha \delta^\beta_{(\mu} \delta^\gamma_{\nu)} 
- \phi^\gamma \delta^\alpha_\mu \delta^\beta_\nu \right),\\ 
\begin{split}
    \tensor{\mathsf{{M}}}{_{\mu\nu}^{\alpha\beta}}&\propto\frac{1}{2A(\phi)^2}\left(g_{\mu\nu}R^{\alpha\beta}+2\tensor{R}{_{(\mu}^\alpha}\delta^{^\beta}_{\nu)}-\delta^{\alpha}_{\mu}\delta^{\beta}_{\nu}R+2\tensor{R}{^\alpha_{\mu\nu}}^{\beta}\right)+\frac{\delta^{\alpha}_{\mu}\delta^{\beta}_{\nu}V(\phi)}{A(\phi)^4}+\\&-\frac{2A'(\phi)}{A(\phi)^3}\left(g_{\mu\nu}\phi^{\beta\alpha}-\delta^{\alpha}_{\mu}\delta^{\beta}_{\nu}\Box\phi\right)-\left(2X\delta^{\alpha}_{\mu}\delta^{\beta}_{\nu}-g_{\mu\nu}\phi^{\alpha\beta}\right)\left(\frac{1}{2}f(\phi)+\frac{6A'(\phi)}{A(\phi)^4}+\frac{2A''(\phi)}{A(\phi)^3}\right),
\end{split}
\end{align}
\end{widetext}
where, the scalar quantity is defined as $X \equiv \nabla_\mu \phi \nabla^\mu \phi/2$, and the following notation is adopted for index symmetrization  $A_{(\mu\nu)} \equiv (A_{\mu\nu} + A_{\nu\mu})/2$.


\bibliographystyle{apsrev4-2}
\bibliography{bibliography.bib}

\end{document}